\documentclass[fleqn,usenatbib,useAMS]{mnras}
\usepackage{graphicx}   
\usepackage{amsmath}    
\usepackage{amssymb}    
\usepackage{multicol}        
\usepackage{bm}         
\usepackage{pdflscape}  
\usepackage[T1]{fontenc}
\usepackage{ae,aecompl}
\usepackage{newtxtext,newtxmath}

\def\aap{A\&A}
\def\apjl{ApJ}
\def\apj{ApJ}
\def\apjs{ApJS}
\def\aj{AJ}
\def\mnras{MNRAS}

\def\pasp{PASP}

\title[GALEX]{A GALEX view of the DA White Dwarf Population}
\author[Wall et al.]
{Renae E. Wall$^1$, Mukremin Kilic$^1$, P. Bergeron$^2$, Nathan D. Leiphart$^1$\\ 
$^1$Homer L. Dodge Department of Physics and Astronomy, University of Oklahoma, 440 W. Brooks St., Norman, OK 73019, USA\\
$^3$D\'epartement de Physique, Universit\'e de Montr\'eal, C.P. 6128, Succ. Centre-Ville, Montr\'eal, QC H3C 3J7, Canada\\
}

\date{\ \ Submitted \today \vspace{-0.5cm}}
\pubyear{2023}

\begin{document}
\label{firstpage}
\pagerange{\pageref{firstpage}--\pageref{lastpage}}
\maketitle

\begin{abstract}

We present a detailed model atmosphere analysis of 14001 DA white dwarfs from the Montreal White Dwarf Database with ultraviolet photometry
from the GALEX mission. We use the 100 pc sample, where the extinction is negligible, to demonstrate
that there are no major systematic differences between the best-fit parameters derived from optical only data and the optical + UV photometry.
GALEX FUV and NUV data improve the statistical errors in the model fits, especially for the hotter white dwarfs with
spectral energy distributions that peak in the UV. Fitting the UV to optical spectral energy distributions also reveals UV-excess or UV-deficit objects.
We use two different methods to identify outliers in our model fits. Known outliers include objects with unusual atmospheric compositions,
strongly magnetic white dwarfs, and binary white dwarfs, including double degenerates and white dwarf + main-sequence systems.
We present a list of 89 newly identified outliers based on GALEX UV data; follow-up observations of these objects will be required to constrain
their nature. Several current and upcoming large scale spectroscopic surveys are targeting $>10^5$ white dwarfs. In addition, the ULTRASAT
mission is planning an all-sky survey in the NUV band. A combination of the UV data from GALEX and ULTRASAT and optical data on these
large samples of spectroscopically confirmed DA white dwarfs will provide an excellent opportunity to identify unusual white dwarfs in
the solar neighborhood.

\end{abstract}


\section{Introduction}

The Galaxy Evolution Explorer (GALEX) is the first space based mission to attempt an all-sky imaging survey in the ultraviolet
\citep[UV,][]{martin05}. In the ten years that it was operational, GALEX surveyed 26000 square degrees of the sky as part of the
all-sky imaging survey in two band passes: Far Ultraviolet (FUV) and Near Ultraviolet (NUV) with central wavelengths of 1528 and 2271 \AA\,
respectively \citep{morrissey05}. Although its primary goal was to study star formation and galaxy evolution, the depth ($m_{\rm AB} \approx 20.5$ mag)
and the large sky coverage of the all-sky imaging survey provide an excellent opportunity to study UV bright objects like hot white dwarfs.

Prior to Gaia, the majority of the white dwarfs in the solar neighborhood were identified through Sloan Digital Sky Survey spectroscopy, which
specifically targeted hot and blue white dwarfs as flux standards \citep[e.g.,][]{kleinman13}. Many of the SDSS white dwarfs have spectral energy
distributions that peak in the UV. Hence, GALEX FUV and NUV data can help constrain the physical parameters of these white dwarfs.
GALEX data will also be useful for cooler white dwarfs; UV photometry will be used to confirm the temperature derived from the optical data, or to
constrain the far red wing of the Lyman $\alpha$ line that dominates the opacity in the blue part of the spectral energy distribution of cool hydrogen
atmosphere white dwarfs \citep{kowalski06}. Yet, GALEX data are under-utilized in the analysis of white dwarfs in the literature, perhaps due to
the relatively strong extinction observed in the UV. 

\citet{wall19} used 1837 DA white dwarfs with high signal to noise ratio spectra and Gaia parallaxes to verify the absolute calibration of the FUV and
NUV data, and refined the linearity corrections derived by \citet{camarota14}. They also empirically derived extinction coefficients for both bands,
finding $R_{\rm FUV} = 8.01$ and $R_{\rm NUV} = 6.72$, where $R$ is the ratio of the total absorption $A_\lambda$ 
to reddening $E(B-V)$ along the line of sight to an object. \citet{wall19} highlighted the utility of their newly derived extinction coefficients
for identifying white dwarfs with unusual UV photometry. By comparing the observed GALEX magnitudes to predictions from the model atmosphere
calculations, they found 12 outliers in the UV, seven of which were previously known, including three double degenerates, two white dwarf + main-sequence
star binaries, one ZZ Ceti, and one double degenerate candidate.

\citet{lajoie07} compared the effective temperatures obtained from the optical and UV spectra of 140 DA white dwarfs from the
{\em IUE} archive. They found that the optical and UV temperatures of the majority of stars cooler than 40000 K and within 75 pc
are in fairly good agreement with $\Delta T_{\rm eff} / T_{\rm optical} \leq10$\%. They also found that the majority of the discrepancies
between the two temperature measurements were caused by interstellar reddening, which affects the UV more than the optical. 
By restricting their analysis to white dwarfs within 75 pc, where the extinction is negligible, they were able to identify several double
degenerate candidates, as well as a DA + M dwarf system, and stars with unusual atmospheric compositions. \citet{lajoie07} thus
demonstrated that unusual white dwarfs can be identified by comparing temperatures derived solely from optical data and UV data.

In this work, we expand the analysis of optical and UV temperature measurements to the DA white dwarfs in the Montreal White Dwarf Database (MWDD)
aided by GALEX UV data and Gaia Data Release 3 astrometry. To identify unusual white dwarfs, we use two methods. First, we compare
the UV and optical temperatures in a manner similar to \citet{lajoie07}. We refer to this as the temperature comparison method. Our second
method follows the analysis of \citet{wall19} and compares the observed and predicted GALEX magnitudes. We refer to this as the magnitude comparison method.

We provide the details of our sample selection in Section 2, the model atmosphere fitting procedure in Section 3, and the results from the
temperature comparison method for the 100 pc sample and the entire MWDD sample in Section 4. Section 5 presents the results from the
magnitude comparison method. We conclude in Section 6.

\section{Sample Selection}

\begin{figure*}
\centering
\includegraphics[width=3in]{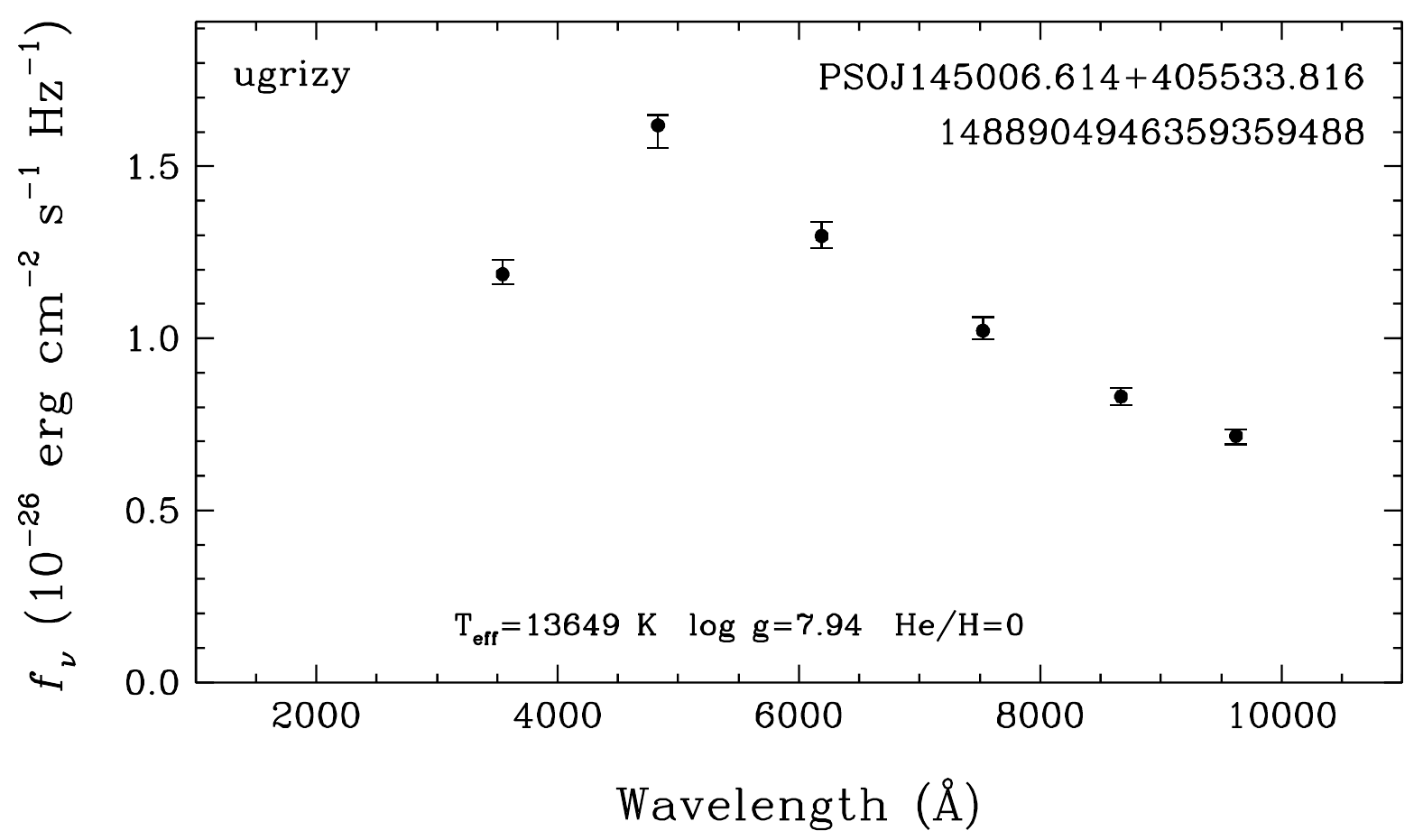}
\includegraphics[width=3in]{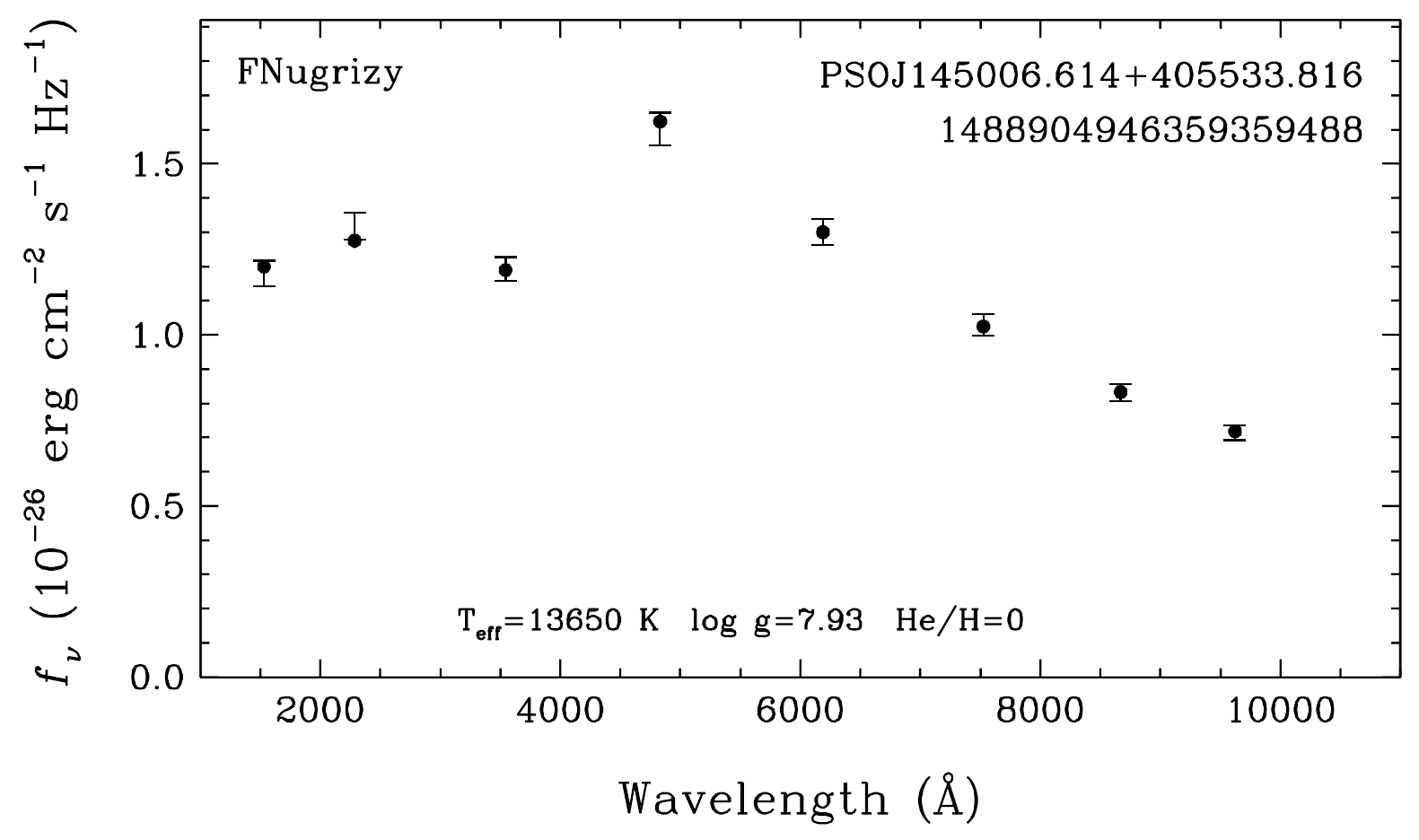}
\includegraphics[width=3in]{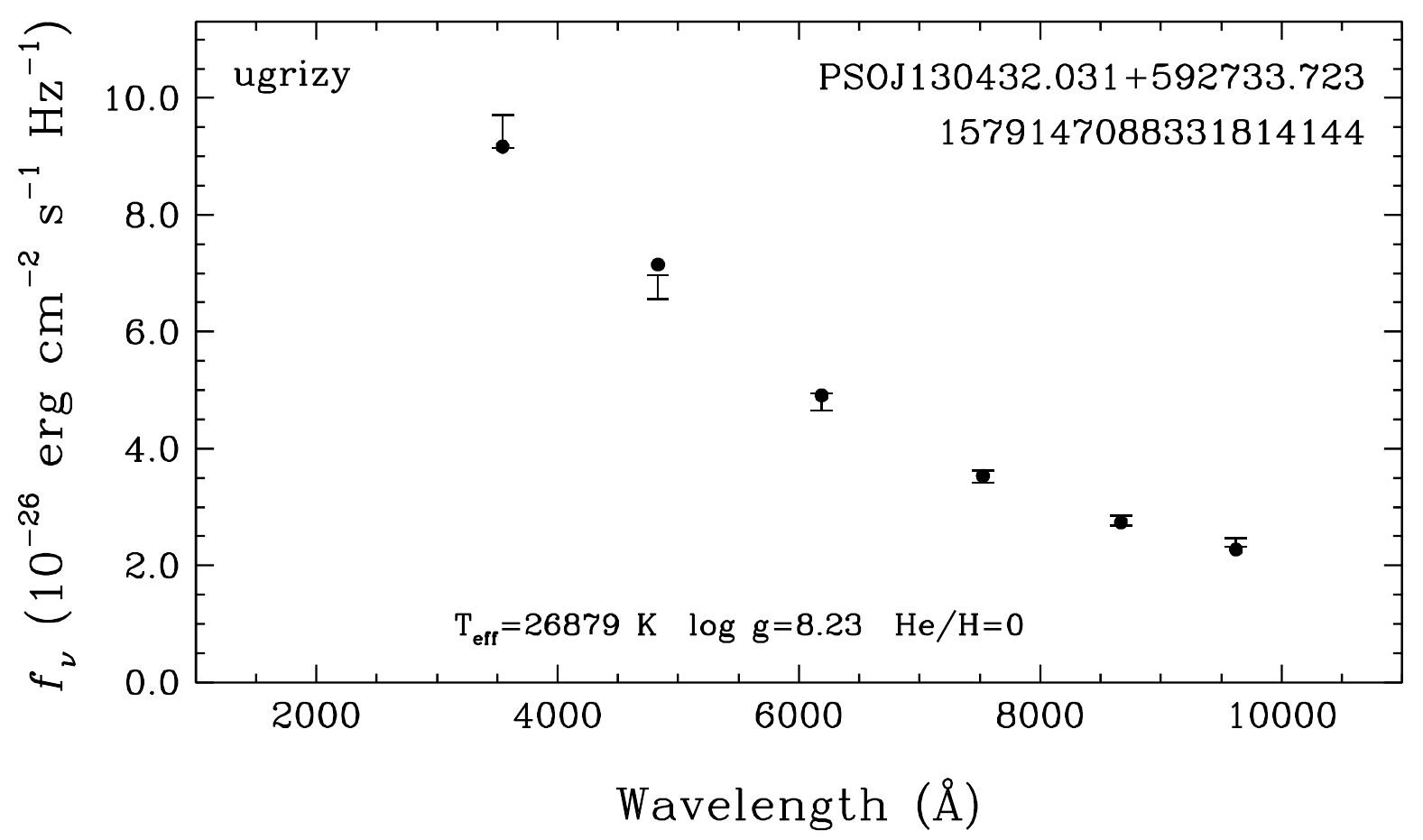}
\includegraphics[width=3in]{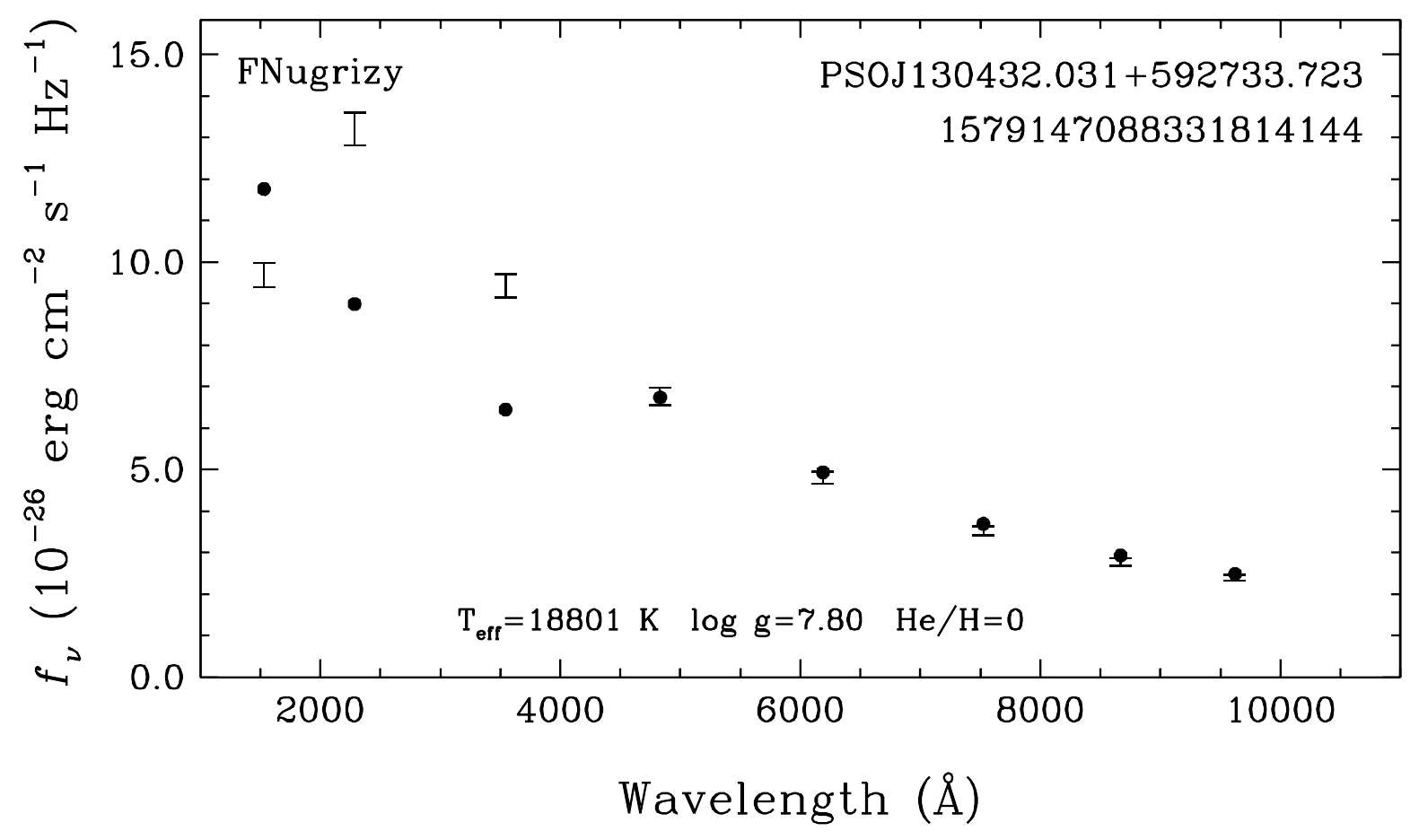}
\caption{{\it Top:} Model fits to WD 1448+411, a spectroscopically confirmed DA white dwarf in the 100 pc SDSS sample.
Each panel shows the best-fitting pure hydrogen atmosphere white dwarf model (filled dots) to the photometry
(error bars). The labels in each panel include the Pan-STARRS coordinates, the Gaia DR3 Source ID, and the
photometry used in the fitting: $FNugrizy$ means GALEX FUV + NUV + SDSS $u$ + Pan-STARRS $grizy$.
The left panel shows the model fits based on the optical data only, whereas the right panel shows the fit using
both optical and the UV data. The best-fitting model parameters are given in each panel.
{\it Bottom:} Model fits to GD 323 (WD 1302+597), a spectroscopically confirmed DAB white dwarf, assuming a pure H atmosphere.
}
\label{figda}
\end{figure*}

We started with all spectroscopically confirmed DA white dwarfs from the Montreal White Dwarf Database \citep{dufour17}
using the September 2022 version of the database. This sample includes over 30000 stars. We removed known white
dwarf + main-sequence binaries and confirmed pulsating white dwarfs from the sample. We then collected the SDSS and Pan-STARRS1 photometry
using the cross-match tables provided by Gaia DR3. We found 25840 DA white dwarfs with Gaia astrometry and Pan-STARRS1
photometry, 20898 of which are also detected in the SDSS.

Gaia DR3 does not provide a cross-matched catalog with GALEX, which performed its all-sky imaging survey between 2003 and 2009.
The reference epoch for the Gaia DR3 positions is 2016. Assuming a 10 year baseline between the GALEX mission and Gaia DR3,
we propagated the Gaia DR3 positions to the GALEX epoch using Gaia proper motions. We then cross-referenced our sample
with the GALEX catalogue of unique UV sources from the all-sky imaging survey (GUVcat) presented in \citet{bianchi17}. We used a cross-match
radius of 3 arcseconds with GUVcat. We found 18456 DA white dwarfs with GALEX data. 

Some of the DA white dwarfs in our sample are bright enough to be saturated in Pan-STARRS, SDSS, or GALEX.
The saturation occurs at $g,r,i \sim 13.5$, $z\sim13$, and $y\sim12$ mag in Pan-STARRS \citep{magnier13}. We remove
objects brighter than these limits. To make sure that there are at least three optical filters available for our model fits,
we limit our sample to objects with at least Pan-STARRS $g,r,i$ photometry available.

We apply the linearity corrections for the GALEX FUV and NUV bands as measured by \citet{wall19}. These corrections
are $\geq 0.5$ mag for FUV and NUV magnitudes brighter than 13th mag. To avoid issues with saturation and large
linearity corrections in the GALEX bands, we further remove objects with FUV and NUV magnitudes brighter than that limit.
We further limit our sample to objects with a $3\sigma$ significant distance measurement \citep{bailer21} so that
we can reliably constrain the radii (and therefore mass and surface gravity) of the stars in our sample. Our final sample contains
14001 DA white dwarfs with photometry in at least one of the GALEX filters and the Pan-STARRS $gri$ filters. 
However, more than half of the stars in our final selection, 7574 of them, have GALEX FUV, NUV, SDSS $u$, and Pan-STARRS
$gri(zy)$ photometry available. 

\section{The Fitting Procedure}

We use the photometric technique as detailed in \citet{bergeron19}, and perform two sets of fits; 1) using
only the optical data, and 2) using both the optical and the UV data. In the first set of fits
we use the SDSS $u$ (if available) along with the Pan-STARRS $grizy$ photometry to model the spectral energy
distribution of each DA white dwarf, and in the second set of fits we add the GALEX FUV (if available) and NUV data.

We correct the SDSS $u$ magnitude to the AB magnitude system using the corrections provided by \citet{eisenstein06}.
For the reasons outlined in \citet{bergeron19}, we adopt a lower limit of 0.03 mag uncertainty in all bandpasses, and
use the de-reddening procedure outlined in \citet{harris06} where the extinction is assumed to be zero for stars
within 100 pc, to be maximum for those located at distances 250 pc away from the Galactic plane, and to vary
linearly along the line of sight between these two regimes.

We convert the observed magnitudes into average fluxes using the appropriate zero points, and compare with the average
synthetic fluxes calculated from pure hydrogen atmosphere models. We define a $\chi^2$ value in terms of the difference
between observed and model fluxes over all bandpasses, properly weighted by the photometric
uncertainties, which is then minimized using the nonlinear least-squares method of Levenberg-Marquardt \citep{press86}
to obtain the best fitting parameters. We obtain the uncertainties of each fitted parameter directly from the covariance
matrix of the fitting algorithm, while we calculate the uncertainties for all other quantities derived from these parameters
by propagating in quadrature the appropriate measurement errors.

We fit for the effective temperature and the solid angle, $\pi (R/D)^2$, where $R$ is the radius of the star and $D$ is its
distance. Since the distance is known from Gaia parallaxes, we constrain the radius of the star directly,
and therefore the mass based on the evolutionary models for white dwarfs. The details of our fitting method, including the
model grids used are further discussed in \citet{bergeron19} and \citet{genest19}. 

\section{Results from Temperature Comparison}

\subsection{The 100 pc SDSS Sample}

We use the 100 pc white dwarf sample in the SDSS footprint to test if the temperatures obtained from the optical and the UV
data agree, and also to test the feasibility of identifying UV-excess
or UV-deficit objects. \citet{kilic20} presented a detailed model atmosphere analysis of the 100 pc white dwarf sample in the
SDSS footprint and identified 1508 DA white dwarfs. Cross-matching this
sample with GUVcat \citep{bianchi17}, we find 847 DA white dwarfs with GALEX data; 377 have both
FUV and NUV photometry available, while 470 have only NUV data available.

The top panels in Figure \ref{figda} show our fits for WD 1448+411, a spectroscopically confirmed DA white dwarf \citep{gianninas11}
in the 100 pc SDSS sample. The top left panel shows the SDSS $u$ and Pan-STARRS $grizy$ photometry (error bars)
along with the predicted fluxes from the best-fitting pure hydrogen atmosphere model (filled dots). The labels in the
same panel give the Pan-STARRS coordinates, Gaia DR3 Source ID, and the photometry used in the fitting. The top right
panel shows the same model fits, but with the addition of the GALEX FUV and NUV photometry. 
The temperature and surface gravity estimates from both sets of fits, based on either the optical data only
(left panel) or a combination of the optical and UV data (right panel), agree remarkably well for this star.
Hence, the spectral energy distribution of WD 1448+411 in the 0.1-1 $\mu$m range is consistent with an
isolated pure hydrogen atmosphere white dwarf.

The bottom panels in Figure \ref{figda} show the model fits for another white dwarf in the 100 pc SDSS sample.
GD 323 (WD 1302+597) is a spectroscopically confirmed DAB white dwarf \citep{wesemael93}.
The use of pure hydrogen atmosphere models to fit its spectral energy distribution is obviously inappropriate.
However, we use GD 323 to demonstrate how fitting the UV to optical spectral energy distribution can
reveal objects with unusual atmospheric composition.  The bottom left panel in Figure \ref{figda} shows our model
fits using only the optical data from the SDSS and Pan-STARRS. Assuming a pure hydrogen composition,
GD 323 would have the best-fitting $T_{\rm eff} = 26879 \pm 1310$ K and $\log{g} = 8.230 \pm 0.047$.
This solution provides an excellent match to the optical photometry. The bottom right panel shows the same model
fits with the addition of the GALEX FUV and NUV data. The best-fitting model parameters are significantly
different, and clearly the pure hydrogen atmosphere models cannot match the UV portion of the spectral
energy distribution of GD 323. Hence, a comparison between the two sets of model fits based on
optical and/or UV data has the potential to identify DAB or other types of unusual objects among the
DA white dwarf population in the solar neighborhood.

\begin{figure}
\centering
\includegraphics[width=3.4in, clip=true, trim=0.3in 2in 0.5in 1.3in]{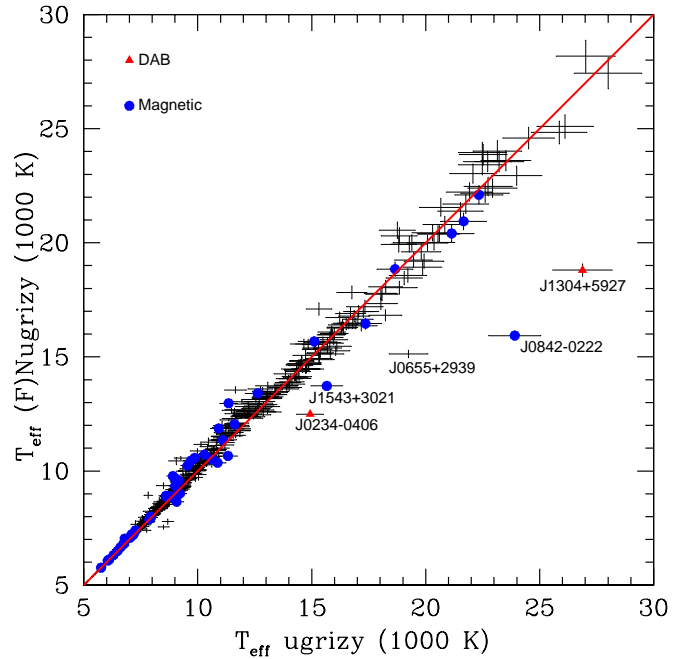}
\caption{A comparison between the effective temperature derived from optical only data versus
a combination of the UV and optical data for the DA white dwarfs in the 100 pc SDSS $\cap$ GALEX sample.
Unusual objects, magnetic DAH and mixed composition DAB white dwarfs, are labeled with blue dots
and red triangles, respectively.}
\label{figsdss}
\end{figure}

\begin{figure}
\centering
\includegraphics[width=3.0in]{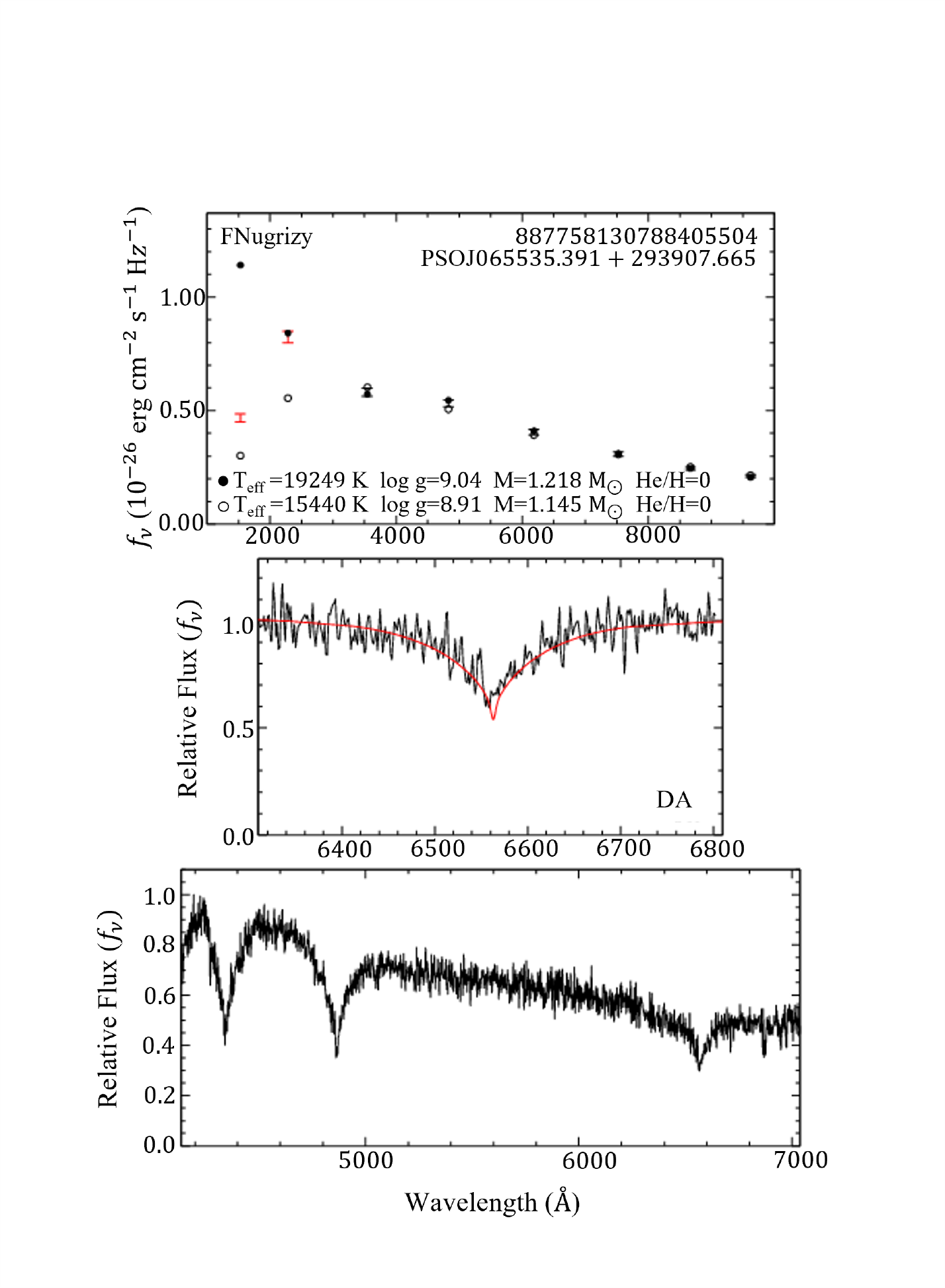}
\caption{Model atmosphere fits to the DA white dwarf J0655+2939. The top panel shows the best-fitting H (filled dots) and He (open circles) atmosphere white dwarf models to the optical photometry (black error bars). The middle panel shows the observed spectrum (black line) along with the predicted spectrum (red line) based on the pure H atmosphere solution. The bottom panel shows a broader wavelength range.}
\label{figdef}
\end{figure}

\begin{figure*}
\centering
\includegraphics[width=3.4in, clip=true, trim=0.3in 2in 0.5in 1.3in]{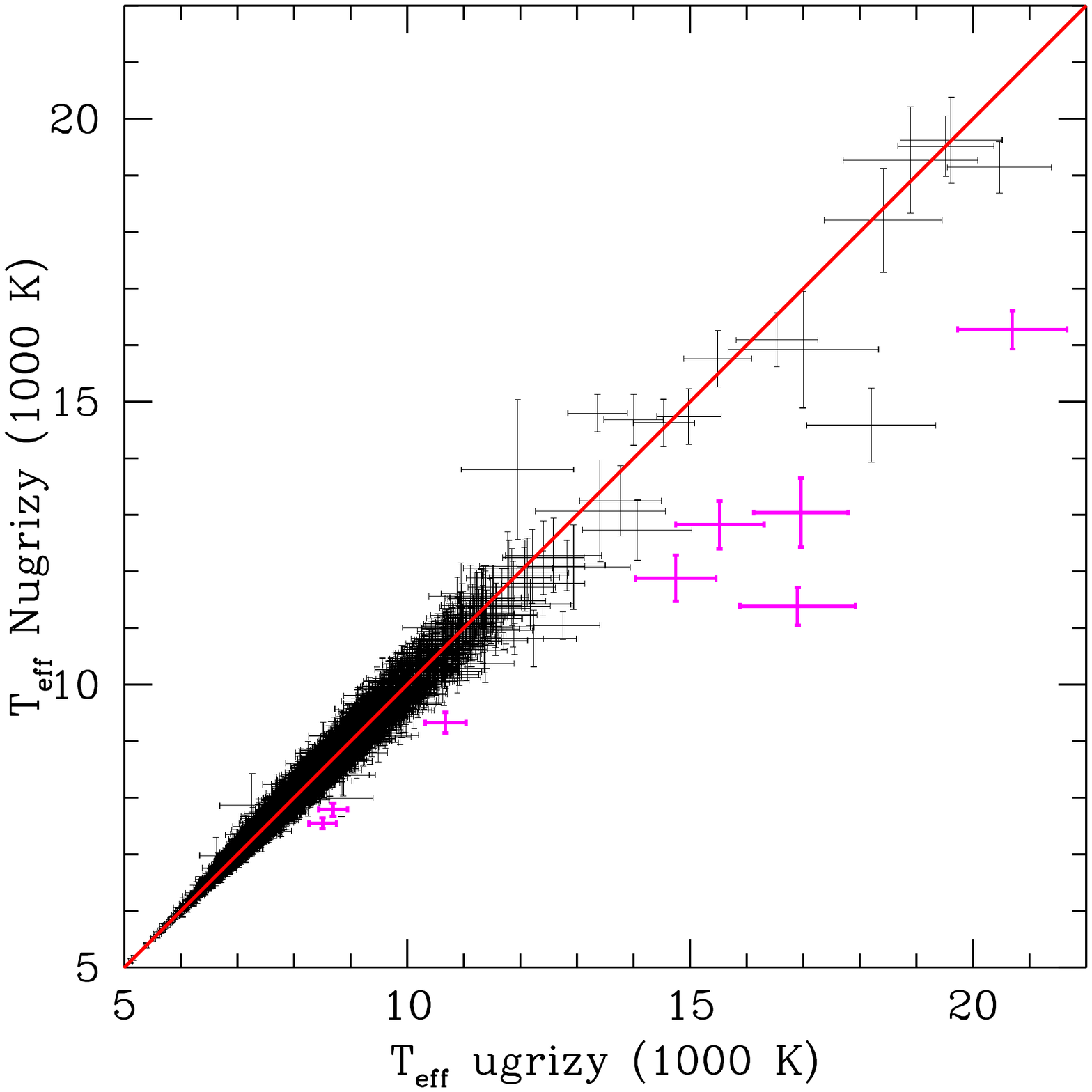}
\includegraphics[width=3.4in, clip=true, trim=0.3in 2in 0.5in 1.3in]{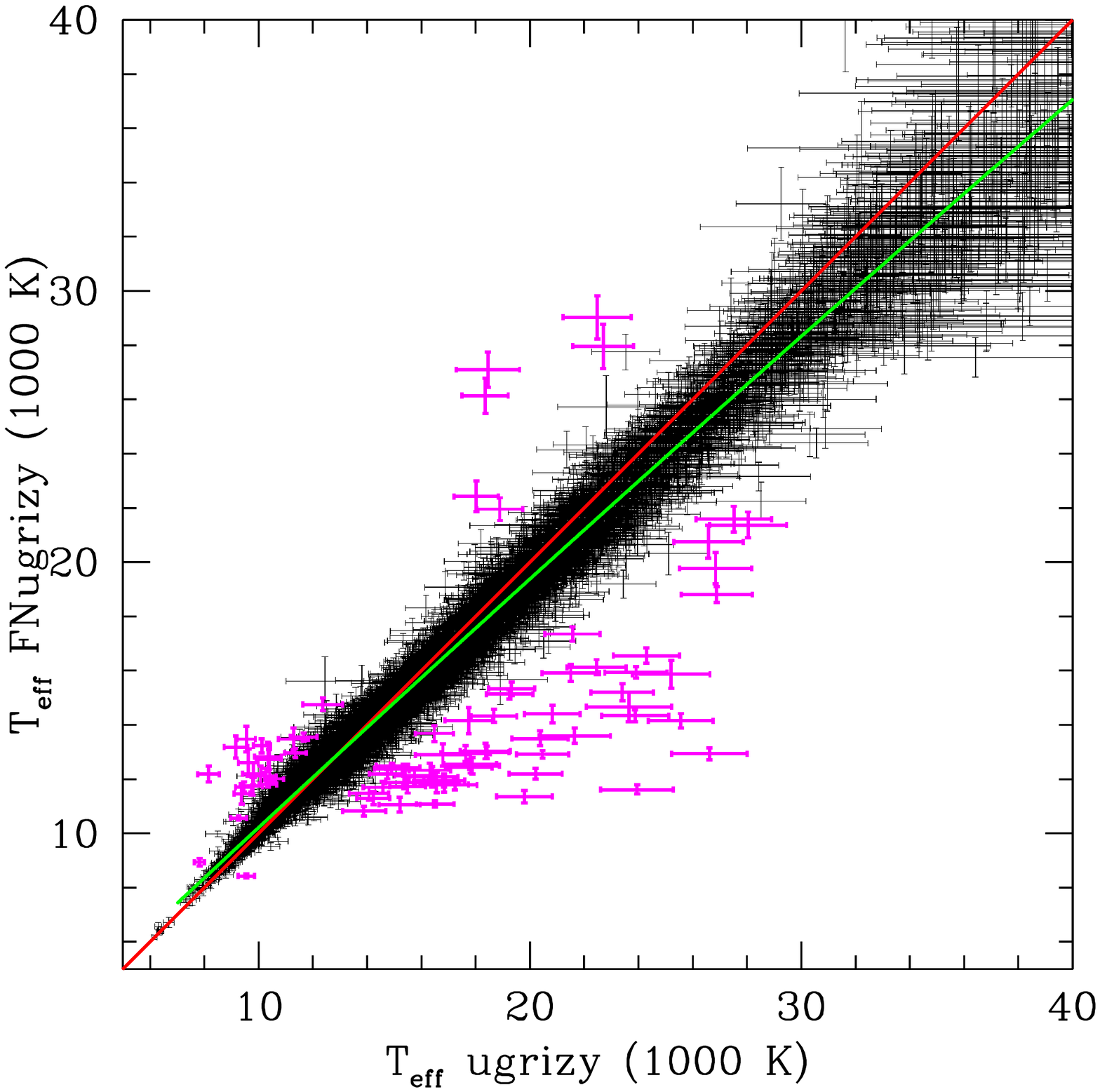}
\caption{A comparison between the effective temperature derived from optical only data (SDSS $u$ and Pan-STARRS
$grizy$) versus the optical + UV data for the DA white dwarfs in the SDSS footprint.
The left panel shows objects with only NUV data, whereas the right panel includes
objects with both FUV and NUV data. The 1:1 line is shown in red. The green line is the best-fitting
polynomial to the data. The $3\sigma$ outliers are shown in magenta.}
\label{figzoomsdss}
\end{figure*}

Figure \ref{figsdss} shows a comparison between the model fits using optical data only versus a combination
of the optical + UV data for the DA white dwarfs in the 100 pc SDSS sample. Blue
dots and red triangles mark the magnetic and DAB white dwarfs, respectively. The majority of the objects in this figure fall
very close to the 1:1 line, shown in red, confirming that they are consistent with pure hydrogen atmosphere
white dwarfs. 

Excluding the five significant outliers labeled in the figure, the effective temperature and $\log{g}$
derived from the GALEX+optical data are slightly higher than the values obtained from the optical data
only by $50^{+215}_{-71}$ K and $0.01^{+0.04}_{-0.01}$ dex, respectively. Hence, there are no major systematic
differences between the best-fit parameters derived from optical only data and the optical + UV photometry.
However, the addition of the GALEX FUV and NUV data helps improve the statistical errors in the model
fits, especially for the hotter white dwarfs where the spectral energy distribution peaks in the UV. For example,
for white dwarfs with $T_{\rm eff}<10000$ K, the statistical errors in optical + UV temperature estimates are on average
better by a factor of 1.3 compared to the errors based on the optical data only, but they are better by a factor
of 2.5 for $T_{\rm eff}>15000$ K. 

The five significant outliers in Figure \ref{figsdss} all appear to be fainter than expected in the UV, and that is why their
best-fitting temperatures based on the optical + UV model fits are cooler than those based on the optical data.
These outliers include two DA white dwarfs with unusual atmospheric composition. J1304+5927
(GD 323, see Figure \ref{figda}) and J0234$-$0406 (PSO J038.5646$-$04.1025). The latter was originally
classified as a DA white dwarf based on a low-resolution spectrum obtained by \citet{kilic20}. Higher signal-to-noise ratio follow-up
spectroscopy by \citet{gentile21b} demonstrated that J0234$-$0406 is in fact a DABZ white dwarf that hosts a gaseous
debris disk. Even though its spectral appearance is visually dominated by broad Balmer absorption lines, the atmosphere
of J0234$-$0406 is actually dominated by helium, and that is why it is an outlier in Figure \ref{figsdss}.

J0842$-$0222 (PSO J130.5623$-$02.3741) and J1543+3021 (PSO J235.8127+30.3595) are both
strongly magnetic and massive white dwarfs with $M>1.1~M_{\odot}$ and unusual optical spectra. 
\citet{schmidt86} noted problems with fitting the UV and optical spectral energy distribution of the strongly magnetic
white dwarf PG 1031+234 with a field stronger than 200 MG. They found that the IUE and optical/infrared fits
cannot be reconciled and that there is no Balmer discontinuity in the spectrum of this object.
They attribute this to the blanketing due to hydrogen lines being grossly different, and the addition of a strong
opacity source (cyclotron absorption). GD 229 is another example of a magnetic white dwarf with
inconsistent UV and optical temperature estimates \citep{green81}. Out of the 51 magnetic
white dwarfs shown in Figure \ref{figsdss}, only J0842$-$0222 and J1543+3021 have significantly discrepant UV and
optical temperatures. Hence, such inconsistencies seem to impact a fraction of the magnetic DA white dwarfs in the solar neighborhood.

Another outlier, J0655+2939 (PSO J103.8966+29.6527), is also a massive white dwarf with $M\sim1.2~M_{\odot}$.
We obtained follow-up optical spectroscopy of J0655+2939 using the KOSMOS spectrograph on the APO 3.5m telescope
on UT 2023 Jan 28. We used the blue grism in the high slit position with a $2.05\arcsec$ slit, providing wavelength coverage from
4150 \AA\ to 7050 \AA\ and a resolution of 1.42 \AA\ per pixel in the $2\times2$ binned mode.

Figure \ref{figdef} shows our model fits for J0655+2939. The top panel shows the best-fitting H (filled dots) and He (open circles)
atmosphere white dwarf models to the optical photometry (black error bars). Note that the GALEX photometry (red error bars) are not used in these fits. The middle panel shows the observed spectrum (black line) along
with the predicted spectrum (red line) based on the pure H atmosphere solution.
The bottom panel shows the entire KOSMOS spectrum. We confirm J0655+2939 as a DA white dwarf. Even though its Balmer
lines and the optical + NUV photometry agree with the pure H atmosphere solution, J0655+2939 is significantly fainter than expected
in the GALEX FUV band. The source of this discrepancy is unclear, but the observed H$\alpha$ line core is also slightly shallower than
expected based on the pure H atmosphere model.

\subsection{The MWDD DA sample}

\begin{figure*}
\centering
\includegraphics[width=3.4in, clip=true, trim=0.3in 2in 0.5in 1.3in]{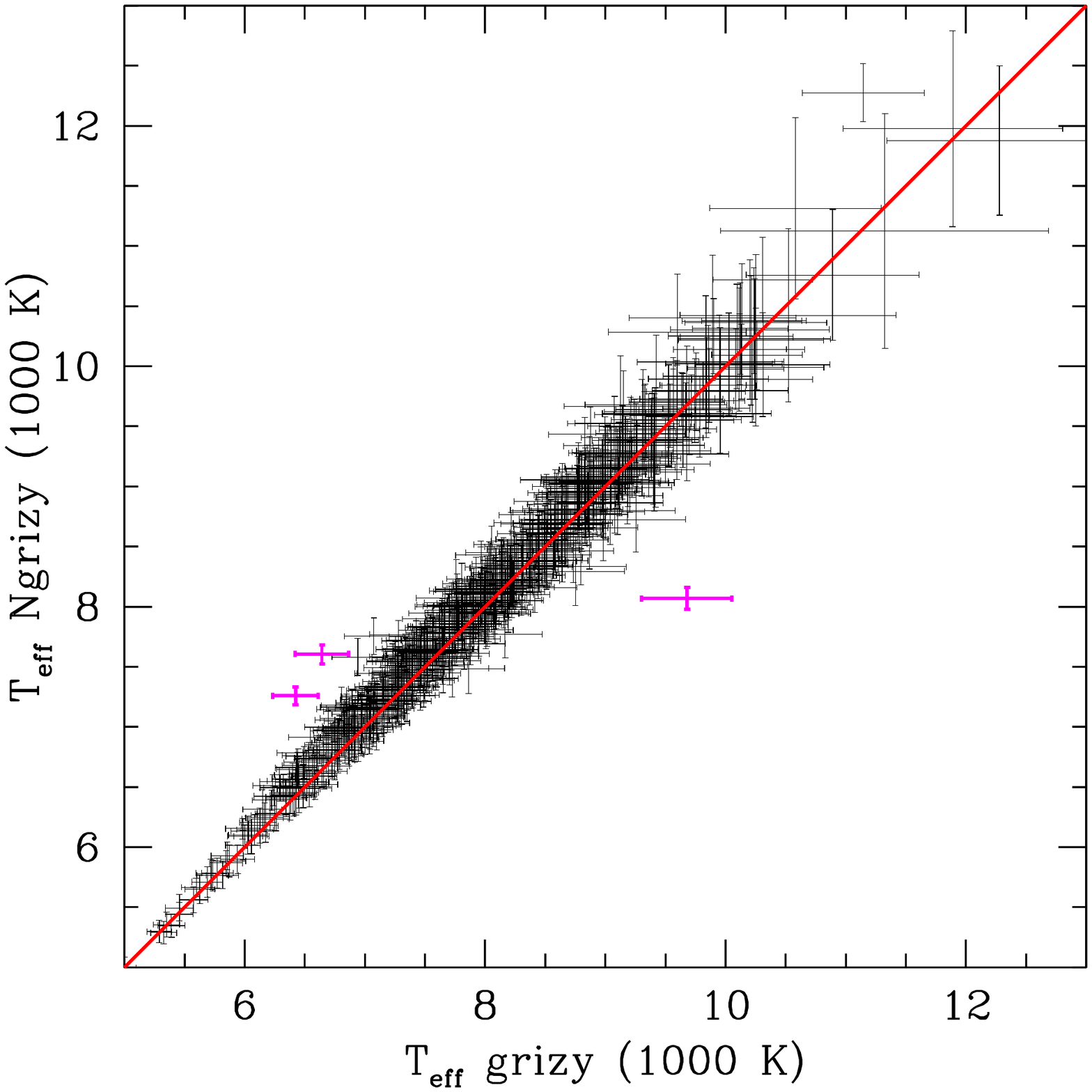}
\includegraphics[width=3.4in, clip=true, trim=0.3in 2in 0.5in 1.3in]{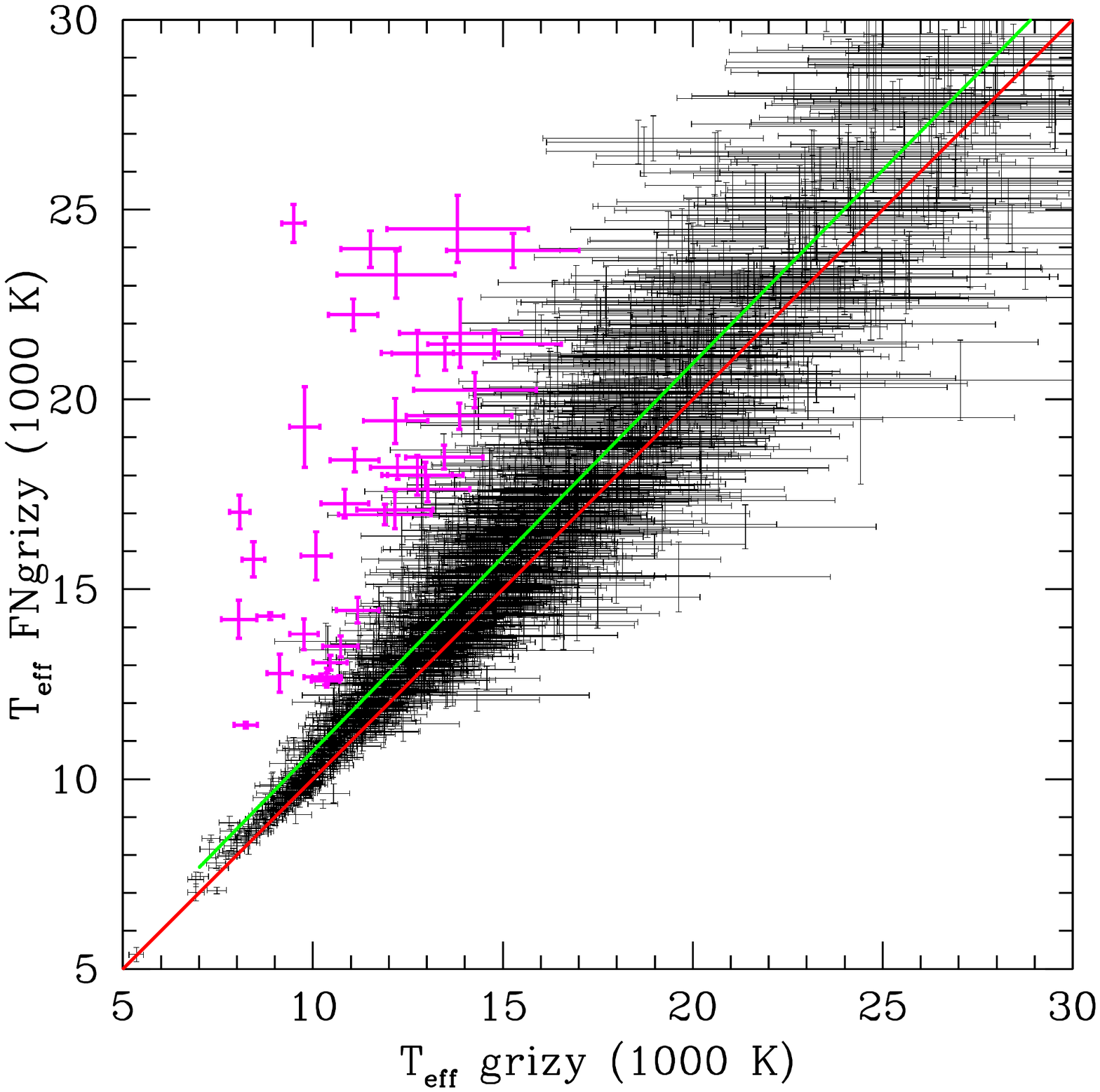}
\caption{Same as Figure \ref{figzoomsdss}, but for the DA white dwarfs outside of the SDSS footprint.}
\label{figzoomps}
\end{figure*}

\begin{figure*}
\centering
\includegraphics[width=3in]{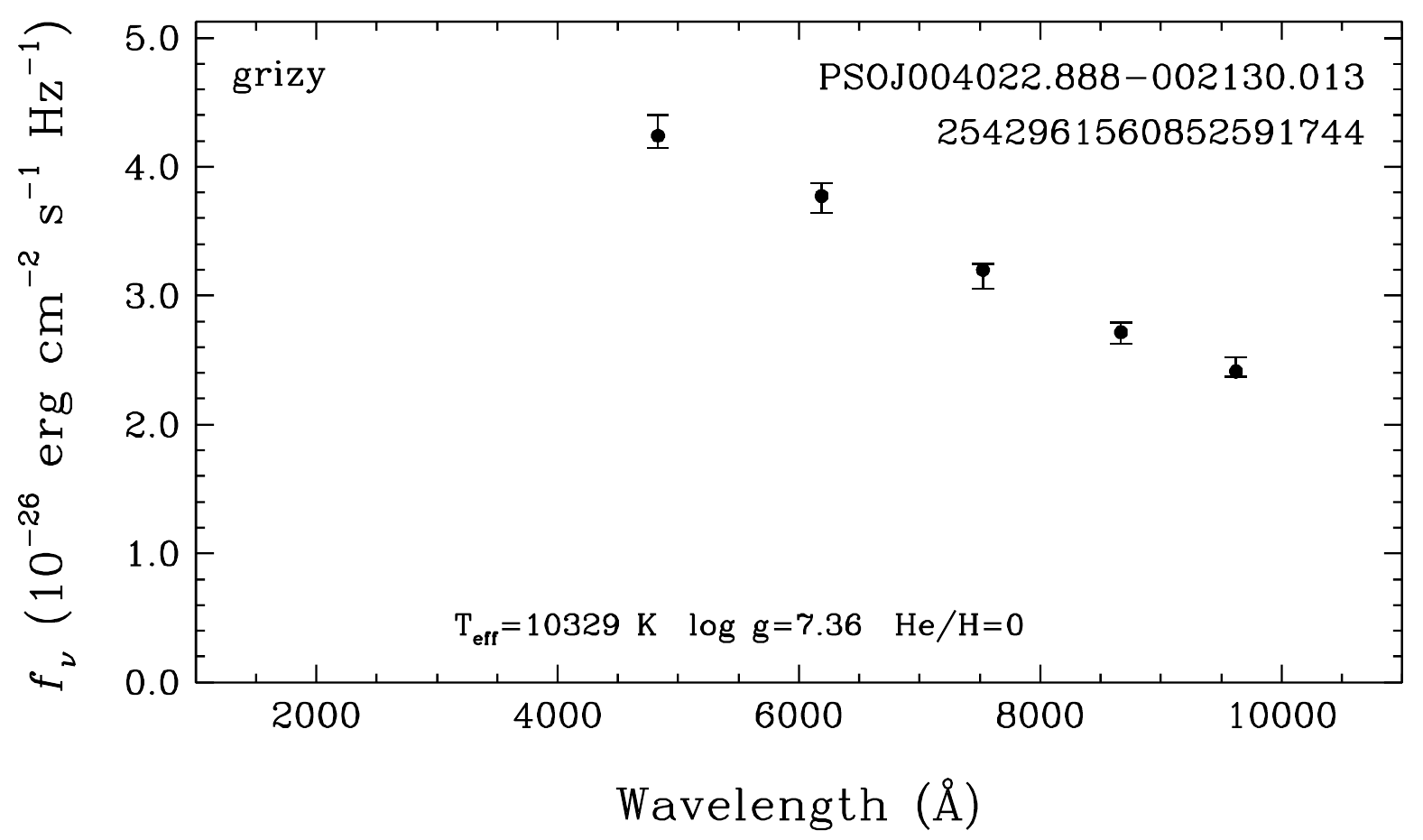}
\includegraphics[width=3in]{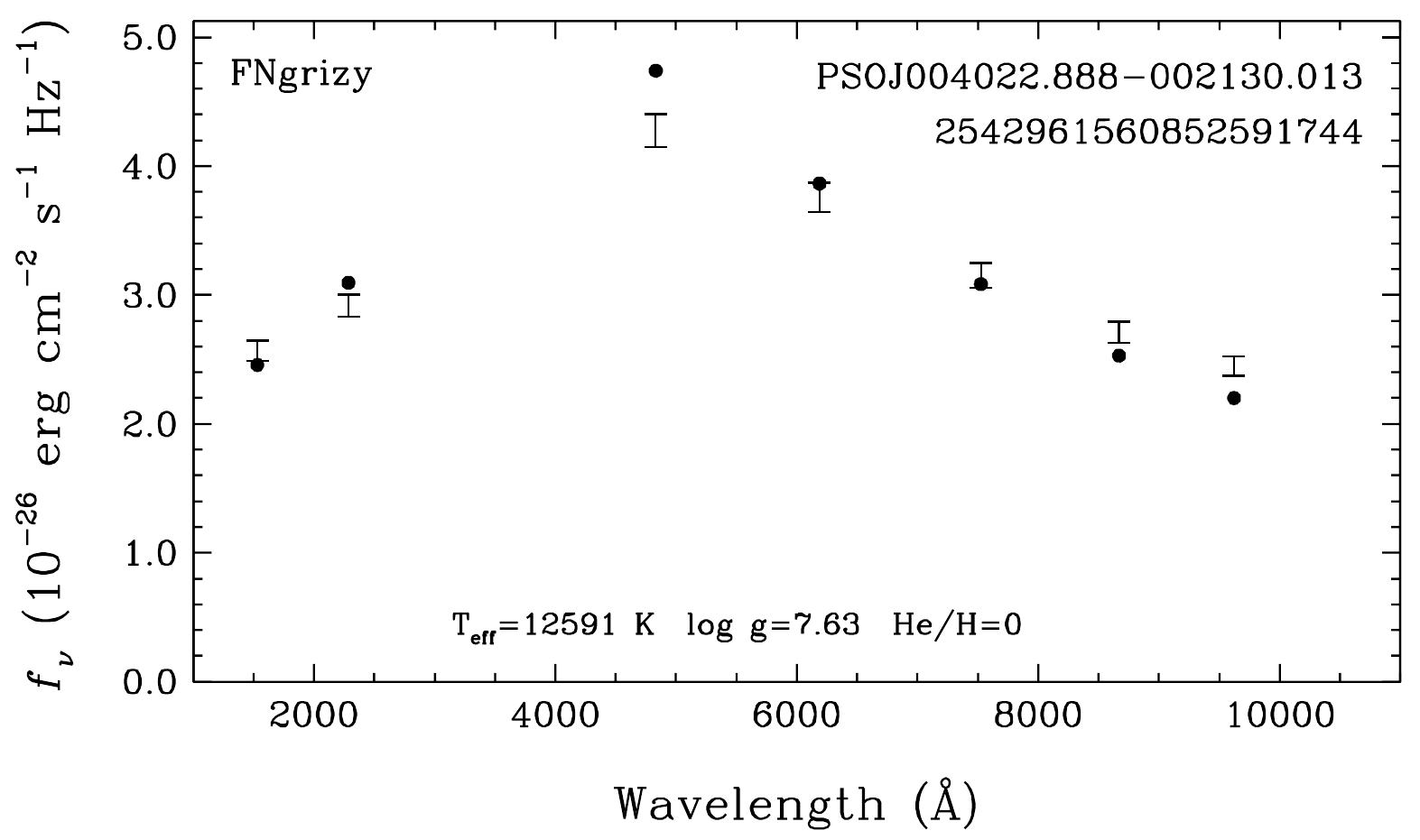}
\includegraphics[width=3in]{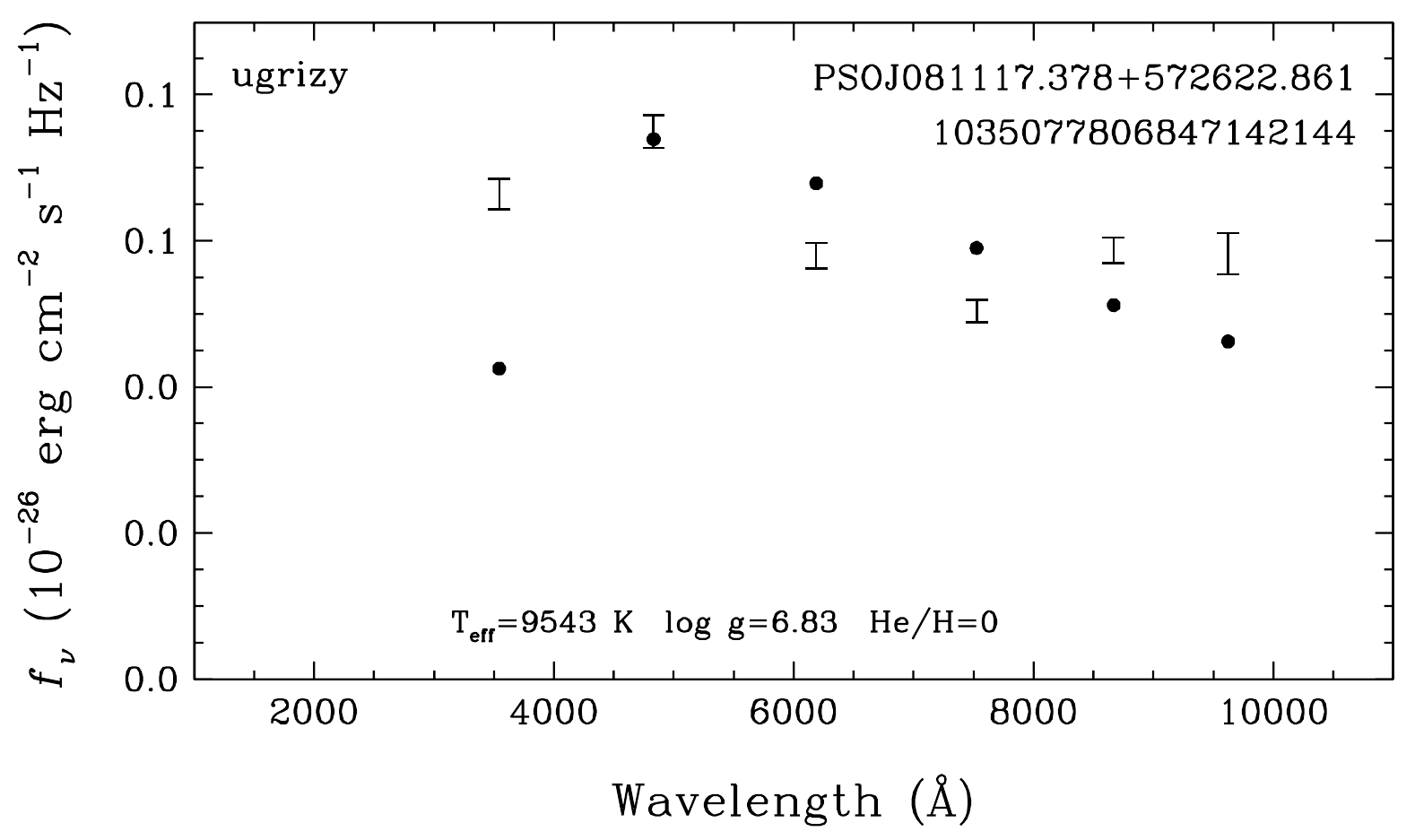}
\includegraphics[width=3in]{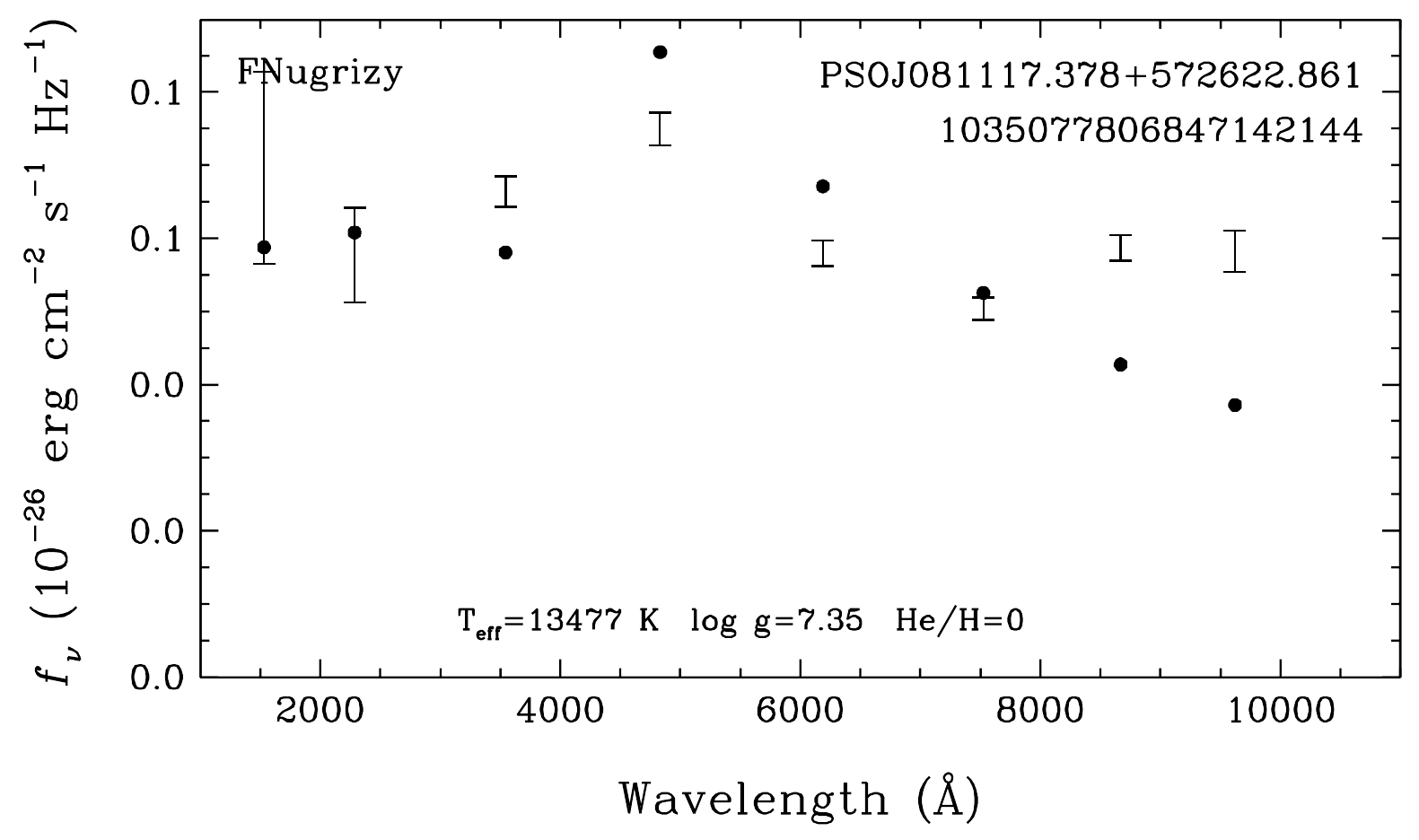}
\caption{Fits to the optical (left) and optical + UV (right) spectral energy distributions of two of the outliers in our
DA white dwarf sample. The top panels show the fits for the double-lined spectroscopic binary WD 0037$-$006,
and the bottom panels show the fits for a previously known DA + M dwarf binary.}
\label{figsbdam}
\end{figure*}

The 100 pc SDSS DA white dwarf sample discussed in the previous section clearly demonstrates that 1) there
are no large-scale systematic differences between the model fits using optical only data ($ugriz$) and a combination
of optical + UV data, and 2) GALEX FUV and NUV data can be used to identify unusual DA white dwarfs with helium-rich
atmospheres or strong magnetic fields. We now expand our study to the entire Montreal White Dwarf Database DA
white dwarf sample in the Pan-STARRS $\cap$ GALEX footprint. 

Figure \ref{figzoomsdss} shows a comparison between the effective temperatures derived from optical and UV data
for the DA white dwarfs in the SDSS footprint. The difference from Figure \ref{figsdss} is that the sample shown here
extends beyond 100 pc, and therefore is corrected for reddening using the de-reddening procedure from \citet{harris06}
and the GALEX extinction coefficients from \citet{wall19}. The left panel includes objects with only NUV data, whereas the
right panel includes objects with both FUV and NUV data. The red line shows the 1:1 line, and the green line
is the best-fitting polynomial to the data. The magenta points mark the outliers that are $3\sigma$ away from both lines. The best-fitting polynomial takes the form
\begin{equation}
y = c_2 x^2 + c_1 x + c_0,
\label{eq:poly}
\end{equation}

\noindent where y is the $T_{\rm eff}FNugrizy/1000$ and x is $T_{\rm eff}ugrizy/1000$. The coefficients are given in table \ref{tab:coeff}.
The sample with the NUV data only (left panel) is limited mostly to white dwarfs with temperatures between 5000 and
12000 K. This is simply an observational bias; hotter white dwarfs would be brighter in the FUV, and therefore they would
have been detected in both NUV and FUV bands. 

A comparison between the model parameters obtained from $ugrizy$ and $Nugrizy$ (left panel) shows that
there are no systematic differences between the two sets of fits. We find eight $3\sigma$ outliers based on this analysis,
all very similar to the outliers shown in Figure \ref{figsdss} with UV flux deficits.

On the other hand, we do find a systematic trend in the temperature measurements from the fits using
the GALEX FUV, NUV, SDSS $u$, and Pan-STARRS $grizy$ filters shown in the right panel. Here the best-fitting
polynomial shows that the temperatures based on the optical + UV data are slightly underestimated compared to
the temperatures obtained from the optical data only. The difference is $-180$ K at 15000 K, $-620$ K at 20000 K, and
$-1670$ K at 30000 K. Note that the average temperature errors based on the optical data are 670, 970, and 1850 K at
15000, 20000, and 30000 K, respectively. Hence, the observed systematic shift in this figure is consistent with the optical
constraints on the same systems within $1\sigma$. We identify 83 outliers $3\sigma$ away from both the 1:1 line
and the best-fitting polynomial (red and green lines in the figure) including a number of UV-excess objects. 

Figure \ref{figzoomps} shows a similar comparison for the DA white dwarfs outside of the SDSS footprint. These
do not have SDSS $u$-band measurements, hence our model fits are based on the Pan-STARRS $grizy$ and
GALEX FUV and NUV bands. The left panel shows the model fits for the DA sample with only NUV data available.
Here the 1:1 line provides an excellent match to the parameters obtained from both the optical and the optical + UV
analysis. We identify only 3 outliers based on this subsample.

The right panel in Figure \ref{figzoomps} reveals a systematic trend in the temperature measurements based on
the GALEX FUV + NUV + $grizy$ data compared to the temperatures derived from the optical only data. 
The best-fitting polynomial takes the form of equation \ref{eq:poly} where y is the $T_{\rm eff}FNgrizy/1000$ and x is $T_{\rm eff}grizy/1000$. The coefficients are given in table \ref{tab:coeff}.
This trend
is similar to the one seen for the SDSS sample (right panel in Figure \ref{figzoomsdss}) but it is in the opposite direction.
The optical + UV analysis leads to temperatures that are slightly over-estimated compared to the analysis using the optical data
only. The difference is +850, +950, and +1090 K at 15000, 20000, and 30000 K, respectively. The average temperature
errors based on the optical data are 670, 2810, and 6040 K at 15000, 20000, and 30000 K, respectively. Again, the observed
systematic trend is consistent with the results from the optical only analysis within $1\sigma$. We identify 41 outliers,
all of which are UV-excess objects, based on this diagram. 

\begin{table*}
\caption{Coefficients for the best-fitting polynomials in figures \ref{figzoomsdss} and \ref{figzoomps}.}
\begin{tabular}{ccc}
\hline
Coefficient & Figure \ref{figzoomsdss} & Figure \ref{figzoomps}\\
$c_0$ & 0.82743420 & 0.48395959\\
$c_1$ & 0.94949536 & 1.02740975\\
$c_2$ & -0.00109481 & -0.00021316\\
\hline
\label{tab:coeff}
\end{tabular}
\end{table*}

\begin{table*}
\small
\caption{The list of outliers that were previously known to be unusual. The horizontal line separates
the UV-deficit (top) and the UV-excess (bottom) objects.}
\begin{tabular}{cccc}
\hline
Object  & Gaia DR3 Source ID &  Spectral Type & Reference\\
\hline
PSO J012.0395$-$01.4109 & 2530629365419780864 &  DA(He) & \citet{kilic20} \\
PSO J017.4701+18.0000 & 2785085218267094784 &  DA(He) & \citet{kepler15} \\
PSO J025.4732+07.7206 & 2571609886069150592 &  DAB & \citet{kepler15} \\
PSO J027.3938+24.0130 & 291186211300158592  &  DZA & \citet{gentile17} \\
PSO J033.0221+06.7391 & 2521035817229538688 &  DA:H: & \citet{kleinman13} \\
PSO J038.5646$-$04.1025 & 2489275328645218560 &  DABZ & \citet{gentile21b}\\
PSO J055.6249+00.4048 & 3263696071424152704 &  DA+DB & \citet{limoges10} \\
PSO J119.5813+35.7453 & 906772187229375104  &  DAH & \citet{kleinman13} \\
PSO J123.8841+21.9779 & 676473944873877248 &  DAB & \citet{kleinman13} \\
PSO J125.6983+12.0296 & 649304840753259520 &  DAH: & \citet{kepler15} \\
PSO J130.5623$-$02.3741 & 3072348715677121280 & DAH?DBH? & \citet{kilic20} \\
PSO J131.8174+48.7057 & 1015028491488955776 & DBH: & \citet{kleinman13} \\
PSO J132.3710+28.9556 & 705246450482748288 &  DAH & \citet{kleinman13} \\
PSO J132.6463+32.1345 & 706974637946866304 &  DABZ & \citet{kong19} \\
PSO J133.2881+58.7267 & 1037873899276147840 &  DABZ & \citet{gentile19} \\
PSO J136.6362+08.1209 & 584319855260594560 &  DAH & \citet{kleinman13} \\
PSO J140.1791+04.8533 & 579476334742123904 &  DA:B:Z: & \citet{kleinman13} \\
PSO J140.7411+13.2557 & 594146225037566976 &  DABZ & \citet{kepler15} \\
PSO J143.7587+44.4946 & 815134799361707392 &  DAH: & \citet{kepler15} \\
PSO J144.9871+37.1739 & 799763528023185280 &  DAB & \citet{kepler15} \\
PSO J150.9846+05.6405 & 3873396705206744064 & DAH & \citet{kleinman13} \\
PSO J154.6449+30.5584 & 742562844335742208 &  DAH & \citet{kleinman13} \\
PSO J179.9671+00.1309 & 3891115064506627840 &  DA(He) & \citet{kilic20} \\
PSO J182.5106+18.0931 & 3949977724441143552 &  DAB & \citet{kepler15} \\
PSO J196.1335+59.4594 & 1579147088331814144 &  DAB & \citet{wesemael93} \\
PSO J198.6769+06.5415 & 3729586288010410496 &  DA(He) & \citet{kepler15} \\
PSO J201.2108+29.5887 & 1462096958792720384 &  DA(He) & \citet{kepler16} \\
PSO J206.1217+21.0809 & 1249447115013660416 &  DABZ & \citet{kleinman13} \\
PSO J211.9610+30.1917 & 1453322271887656448 &  DA:H: & \citet{kleinman13} \\
PSO J218.8923+04.5738 & 3668901977825959040 &  DAX & \citet{kepler15} \\
PSO J223.2567+06.8724 & 1160931721694284416 &  DA:H: & \citet{kleinman13} \\
PSO J223.9933+18.2145 & 1188753901361576064 &  DA:H: & \citet{kleinman13} \\
PSO J234.3569+51.8575 & 1595298501827000960 &  DBA & \citet{kleinman13} \\
PSO J240.2518+04.7101 & 4425676551115360512 &  DAH & \citet{kleinman13} \\
PSO J261.1339+32.5709 & 1333808965722096000 &  DAH & \citet{kepler15} \\
PSO J341.2484+33.1715 & 1890785517284104960 &  DAH/DQ & \citet{kepler16} \\
PSO J356.5226+38.8938 & 1919346461391649152 &  DAH & \citet{kleinman13} \\
\hline
PSO J010.0954$-$00.3584 & 2542961560852591744 &  DA+DA & \citet{napiwotzki20} \\ 
PSO J042.5074$-$04.6175 & 5184589747536175104 &  DAH: & \citet{kepler16} \\
PSO J051.5805+13.5189 & 17709047809907584   &  DAH & \citet{kilic20} \\
PSO J063.1211$-$11.5012 & 3189613692364776576 &  DA+DA & \citet{napiwotzki20} \\
PSO J065.0980+47.5929 & 257933852944165120  &  DAB & \citet{verbeek12} \\
PSO J094.8914+55.6121 & 997854527884948992  &  DAO & \citet{gianninas11} \\
PSO J109.2922+74.0109 & 1112171030998592256 &  DAM & \citet{marsh96}\\
PSO J122.8223+57.4396 & 1035077806847142144 &  DAM & \citet{rebassa16} \\
PSO J123.9537+47.6772 & 931238043230275968 &  DAM & \citet{farihi10} \\
PSO J140.2868+13.0199 & 594229753561550208 &  DAH & \citet{kleinman13} \\
PSO J173.7025+46.8094 & 785521450828261632 &  DD? & \citet{bedard17} \\
PSO J182.0967+06.1655 & 3895444662122848512 &  DAM & \citet{rebassa16} \\
PSO J224.1602+10.6747 & 1180256944222072704 &  DAM & \citet{rebassa16} \\
PSO J337.4922+30.4024 & 1900545847646195840 &  DAM? & \citet{rebassa19} \\
PSO J344.9451+16.4879 & 2828888597582293760 &  DAM & \citet{farihi10} \\
\hline
\label{outknown}
\end{tabular}
\end{table*} 

In total we identify 135 outliers based on this analysis. Because the full width at half-maximum of the GALEX point
spread function is about 5 arcsec \citep{morrissey07}, blending and contamination from background sources is an
issue. We checked the Pan-STARRS stacked images for each of these outliers to identify nearby
sources that could impact GALEX, SDSS, or Pan-STARRS photometry measurements. We found that 24 of these
outliers were likely impacted by blending sources, reducing the final sample size to 111 outliers. 

Table \ref{outknown} presents the list of 52 outliers that were previously known to be unusual. This list includes
four objects that are confirmed or suspected to be double white dwarfs (PSO J010.0954$-$00.3584,
J055.6249+00.4048,  J063.1211$-$11.5012, and J173.7025+46.8094), 20 confirmed
or suspected magnetic white dwarfs, seven DA + M dwarf systems, and 21 objects with an unusual atmospheric
composition (DAB etc).  

Figure \ref{figsbdam} shows the spectral energy distributions for two of these outliers. The top panels show the fits to the optical and UV + optical
spectral energy distributions of the previously known double-lined spectroscopy binary WD 0037$-$006 \citep{napiwotzki20}.
Under the assumption of a single star, the Pan-STARRS photometry for WD 0037$-$006 indicates
$T_{\rm eff} = 10330 \pm 380$ K and $\log{g} = 7.36 \pm 0.05$. Adding the GALEX FUV and NUV data, the best-fitting solution
significantly changes to $T_{\rm eff} = 12590 \pm 100$ K and $\log{g} = 7.63 \pm 0.01$. In addition, this solution has
problems matching the entire spectral energy distribution, indicating that there is likely a cooler companion contributing significant
flux. This figure demonstrates that double-lined spectroscopic binaries with significant temperature differences between the
primary and the secondary star could be identified based on an analysis similar to the one presented here. 
A similar and complementary method for identifying double-lined spectroscopic binaries was pioneered by \citet{bedard17},
which use optical photometry and spectroscopy to identify systems with inconsistent photometric and spectroscopic solutions.

The bottom panels in Figure  \ref{figsbdam} show the fits to a previously confirmed DA + M dwarf system in our sample
\citep{rebassa16}. Here the optical data is clearly at odds with a single DA white dwarf, and GALEX FUV and NUV
data reveal UV-excess from a hotter white dwarf. The analysis using $FNugrizy$ photometry confirms excess
emission in the Pan-STARRS $zy$-bands, consistent with an M dwarf companion.

\begin{table*}
\centering
\caption{Newly identified outliers among the DA white dwarf population with GALEX data. The horizontal line separates
the UV-deficit (top) and the UV-excess (bottom) objects. }
\begin{tabular}{crrrlll}
\hline
Object  & Gaia DR3 Source ID &  Optical  & Optical + UV  & Spectral & Reference & Notes\\
  &  & $T_{\rm eff}$ (K) & $T_{\rm eff}$ (K) & Type & & \\
\hline
PSO J018.6848+35.4095 & 321093335597030400  & 15369 $\pm$  678 & 11872 $\pm$ 223 &  DA & \citet{gentile15} &  DA(He) LAMOST \\
PSO J032.2011+12.2256 & 73623921366683008   & 27516 $\pm$ 1379 & 21586 $\pm$ 476 &  DA & \citet{kleinman13} &  \\
PSO J043.8655+02.6202 & 1559111783825792    &  8685 $\pm$  254 &  7788 $\pm$ 117 &  DA & \citet{kilic20} &  \\
PSO J056.0479+15.1626 & 42871199614383616   &  8503 $\pm$  241 &  7548 $\pm$  90 &  DA & \citet{andrews15} &  \\
PSO J103.8966+29.6527 & 887758130788405504  & 19249 $\pm$  849 & 15130 $\pm$ 168 &  DA & \citet{kilic20} &  massive \\
PSO J130.7484+10.6677 & 598412403168328960  & 15748 $\pm$  731 & 12139 $\pm$ 293 & DAZ & \citet{kepler15} &  DZA SDSS\\
PSO J132.2963+14.4454 & 608922974120358784  & 19793 $\pm$ 1038 & 11354 $\pm$ 226 &  DA & \citet{gentile19} &   \\
PSO J139.0499+34.9872 & 714469355877947136  & 23646 $\pm$ 1571 & 14667 $\pm$ 545 &  DA & \citet{kleinman13} & massive \\
PSO J151.1401+40.2417 & 803693216941983232  & 13881 $\pm$  797 & 10816 $\pm$ 188 &  DA & \citet{kepler15} &  DA(He) SDSS \\
PSO J158.8293+27.2510 & 728222390915647872  & 17876 $\pm$  998 & 12484 $\pm$ 307 &  DA & \citet{gentile19} &   \\
PSO J159.5929+37.3533 & 751930335511863040  & 14891 $\pm$  628 & 12383 $\pm$ 194 &DA:DC: & \citet{kleinman13} &  DAB SDSS \\
PSO J163.5943$-$02.7860 & 3801901270848297600 & 25209 $\pm$ 1429 & 15858 $\pm$ 522 &  DA & \citet{croom01} &  massive\\
PSO J172.6518$-$00.3655 & 3797201653208863360 & 15198 $\pm$  761 & 11058 $\pm$ 264 &DA:Z & \citet{kleinman13} &  DZA SDSS  \\
PSO J180.6015+40.5822 & 4034928775942285184 & 16551 $\pm$  835 & 11880 $\pm$ 377 &  DA & \citet{kleinman13} &  DC: SDSS \\
PSO J196.7725+49.1045 & 1554826818838504576 & 14573 $\pm$  736 & 11689 $\pm$ 225 & DA: & \citet{kleinman13} &  DA(He) SDSS \\
PSO J213.8277+31.9308 & 1477633195532154752 & 16960 $\pm$  829 & 13040 $\pm$ 608 & DAZ & \citet{gentile19} &  DZA SDSS \\
PSO J215.2971+38.9912 & 1484931581918492544 & 17743 $\pm$  873 & 14159 $\pm$ 483 & DA: & \citet{kleinman13} &  DC: SDSS \\
PSO J231.8495+06.7581 & 1162614902197098624 & 16792 $\pm$ 1013 & 12897 $\pm$ 411 &  DA & \citet{carter13} &  massive \\
PSO J249.3471+53.9644 & 1426634650780861184 & 16315 $\pm$  761 & 12321 $\pm$ 257 & DAZ & \citet{kepler16} &  \\
PSO J309.1036+77.8178 & 2290767158609770240 & 28040 $\pm$ 1427 & 21372 $\pm$ 486 &  DA & \citet{bedard20} &  \\
PSO J324.1725+01.0846 & 2688259922223271296 & 16404 $\pm$  937 & 12078 $\pm$ 389 &  DA & \citet{vidrih07} &  \\
PSO J338.5445+25.1894 & 1877374842678152704 & 16838 $\pm$  892 & 11787 $\pm$ 266 &  DA & \citet{gentile19} &  DA(He) SDSS \\
PSO J342.5363+22.7580 & 2836800855054851456 & 26580 $\pm$ 1279 & 20752 $\pm$ 605 &  DA & \citet{bedard20} &  \\
PSO J348.7601+22.1674 & 2838958711048617856 & 26837 $\pm$ 1327 & 19773 $\pm$ 571 &  DA & \citet{kleinman13} &  \\
\hline
PSO J003.9449$-$30.1015 & 2320237751020937728 &  9768 $\pm$  375 & 13816 $\pm$ 408 &  DA & \citet{vennes02} &  \\
PSO J009.0492$-$17.5443 & 2364297204875140224 & 13479 $\pm$ 1408 & 21203 $\pm$ 437 &  DA & \citet{gianninas11} &  \\
PSO J015.0435$-$28.1077 & 5033974938207807488 & 13023 $\pm$ 1105 & 17630 $\pm$ 337 &  DA & \citet{croom04} &  \\
PSO J019.3103+24.6726 & 294062563782633216  & 12160 $\pm$ 1007 & 17090 $\pm$ 490 &  DA & \citet{kleinman13} &  \\
PSO J021.9568+73.4798 & 535482641132742400  &  6422 $\pm$  190 &  7259 $\pm$  75 &  DA & \citet{limoges15} &  \\
PSO J023.0575$-$28.1766 & 5035296654263954304 & 12745 $\pm$  931 & 17999 $\pm$ 527 &  DA & \citet{croom04} &  \\
PSO J029.9572$-$27.8589 & 5024390701506507648 & 11176 $\pm$  562 & 14449 $\pm$ 337 &  DA & \citet{croom04} &  \\
PSO J041.4724$-$12.7058 & 5158731712247303040 &  9493 $\pm$  307 & 24634 $\pm$ 504 &  DA & \citet{kilkenny16} &  DAM? \\
PSO J051.6792+69.4045 & 494644717692834944  & 13855 $\pm$ 1393 & 19565 $\pm$ 343 &  DA & \citet{gianninas11} &  \\
PSO J052.0294+52.9603 & 443375555640546944  & 10729 $\pm$  467 & 13492 $\pm$ 282 &  DA & \citet{verbeek12} &  \\
PSO J052.2834+52.7335 & 443274778529615232  & 10363 $\pm$  382 & 12680 $\pm$ 249 &  DA & \citet{verbeek12} &  \\
PSO J102.2271+38.4434 & 944388335442133888  & 13883 $\pm$ 1606 & 21741 $\pm$ 901 &  DA & \citet{kleinman13} &  \\
PSO J125.7399+57.8364 & 1034975243028553600 & 18178 $\pm$ 2420 & 31076 $\pm$ 1012 & DA & \citet{bedard20} &  \\
PSO J143.4929+17.7146 & 632864633657062400  & 13811 $\pm$ 1864 & 24486 $\pm$ 883 &  DA & \citet{bedard20} &  DAM? \\
PSO J143.5436+22.4702 & 644043544469790720  & 14268 $\pm$ 1613 & 20245 $\pm$ 465 &  DA & \citet{kleinman13} &  DAM? \\
PSO J146.2852+62.7948 & 1063508669280315776 &  9623 $\pm$  370 & 12604 $\pm$ 502 &  DA & \citet{kleinman13} &  DAM? \\
PSO J149.4751+85.4946 & 1147853241336105344 & 28499 $\pm$ 4417 & 50953 $\pm$ 4975 & DA & \citet{gianninas11} &  \\
PSO J150.3866+01.5162 & 3835962526168788608 & 22704 $\pm$ 1127 & 27966 $\pm$ 821 &  DA & \citet{kepler15} &  \\
PSO J167.1417+31.8979 & 757803896562843392  & 14778 $\pm$ 1760 & 21458 $\pm$ 382 &  DA & \citet{gianninas11} &  \\
PSO J192.2894+24.0266 & 3957635410611476096 & 10266 $\pm$  381 & 12108 $\pm$ 291 &  DA & \citet{kleinman13} &  \\
PSO J200.6206+01.0147 & 3688065808367722368 &  9383 $\pm$  291 & 11459 $\pm$ 371 &  DA & \citet{croom04} &  DAM? \\
PSO J211.4189+74.6498 & 1712016196599965312 &  8237 $\pm$  305 & 11420 $\pm$  77 &  DA & \citet{mickaelian08} &  resolved DAM\\
PSO J218.2047+01.7710 & 3655853106971493760 & 10341 $\pm$  329 & 11793 $\pm$  97 &  DA & \citet{kleinman13} &  \\
PSO J221.4238+41.2449 & 1489712503290614912 &  9786 $\pm$  396 & 19274 $\pm$ 1067 & DA & \citet{bedard20} &  DAM? \\
PSO J223.4269+46.9171 & 1590342178286505216 & 12174 $\pm$  851 & 19440 $\pm$ 593 &  DA & \citet{kleinman13} &  \\
PSO J240.6992+43.8100 & 1384551977098980608 & 10070 $\pm$  352 & 12113 $\pm$ 369 &  DA & \citet{kleinman13} &  resolved DAM? \\
PSO J240.8660+19.6618 & 1203265358904378880 & 10085 $\pm$  401 & 15880 $\pm$ 632 &  DA & \citet{kleinman13} &  DAM? \\
PSO J244.6129+20.5911 & 1202035422006406400 &  9808 $\pm$  394 & 12191 $\pm$ 410 &  DA & \citet{kleinman13} &  resolved DAM? \\
PSO J259.6449+01.9471 & 4387171623850187648 & 10215 $\pm$  438 & 12685 $\pm$  84 &  DA & \citet{mccleery20} &   \\
PSO J263.1394+32.8366 & 4601788317833882240 &  9790 $\pm$  353 & 11747 $\pm$ 336 &  DA & \citet{kepler15} &  DAM? \\
PSO J276.0344+35.2718 & 2095603539740855296 & 11063 $\pm$  653 & 22229 $\pm$ 415 &  DA & \citet{mickaelian08} &  DAM? \\
PSO J334.7157$-$29.4534 & 6615258025441899776 & 10838 $\pm$  624 & 17254 $\pm$ 376 &  DA & \citet{croom04} &  DAM? \\
PSO J346.5586$-$28.0099 & 6606686198432918656 & 11516 $\pm$  779 & 23961 $\pm$ 485 &  DA & \citet{croom04} &  resolved DAM? \\
PSO J352.1333$-$30.0610 & 2329285662270302976 & 12190 $\pm$ 1555 & 23284 $\pm$ 616 &  DA & \citet{vennes02} &  \\
PSO J355.9551+38.5749 & 1919325605029184000 & 10573 $\pm$  344 & 12004 $\pm$  78 &  DA & \citet{kleinman13} &  \\
\hline
\label{outnew}
\end{tabular}
\end{table*}

\begin{figure*}
\centering
\includegraphics[width=2.3in]{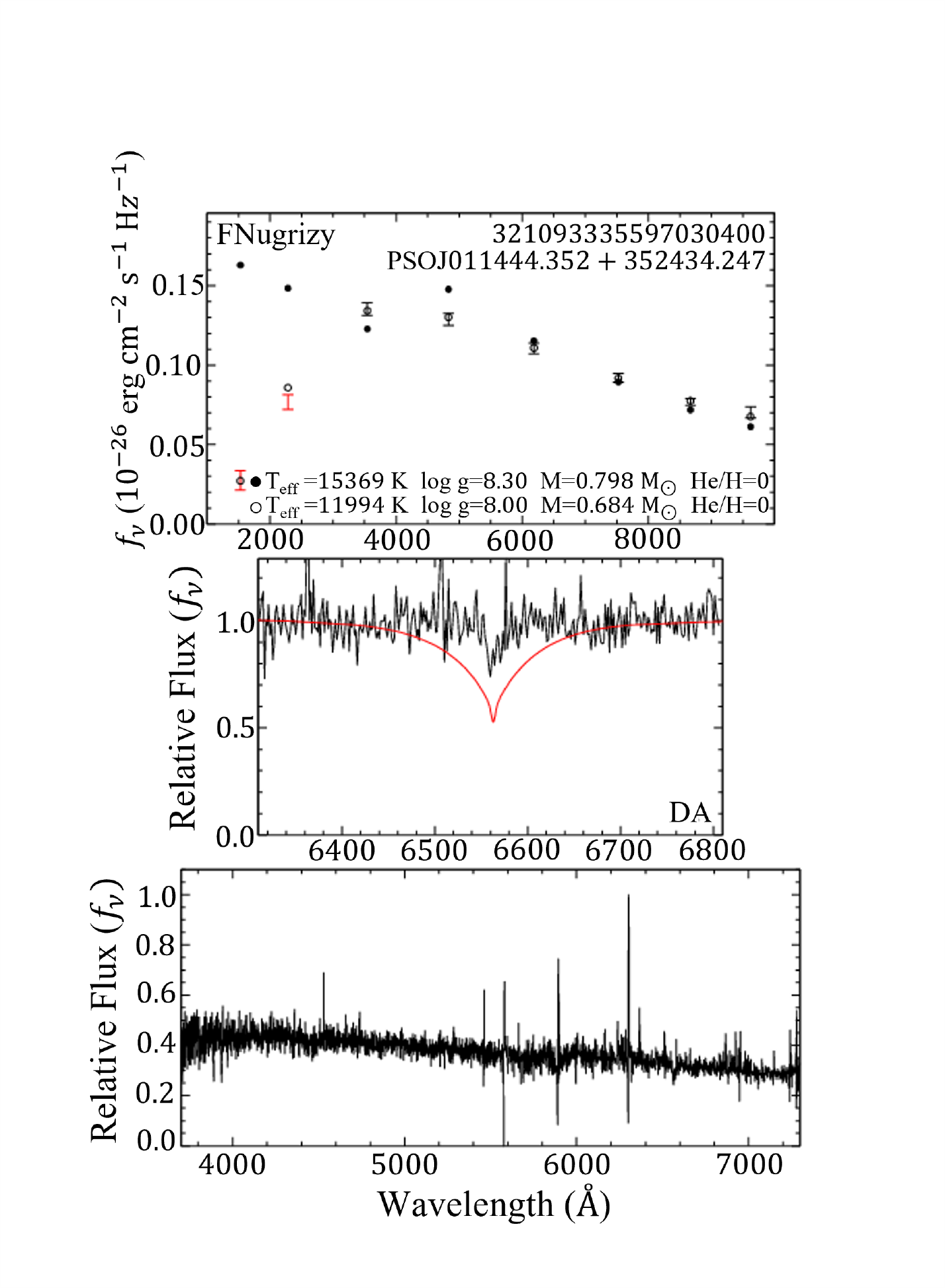}
\includegraphics[width=2.3in]{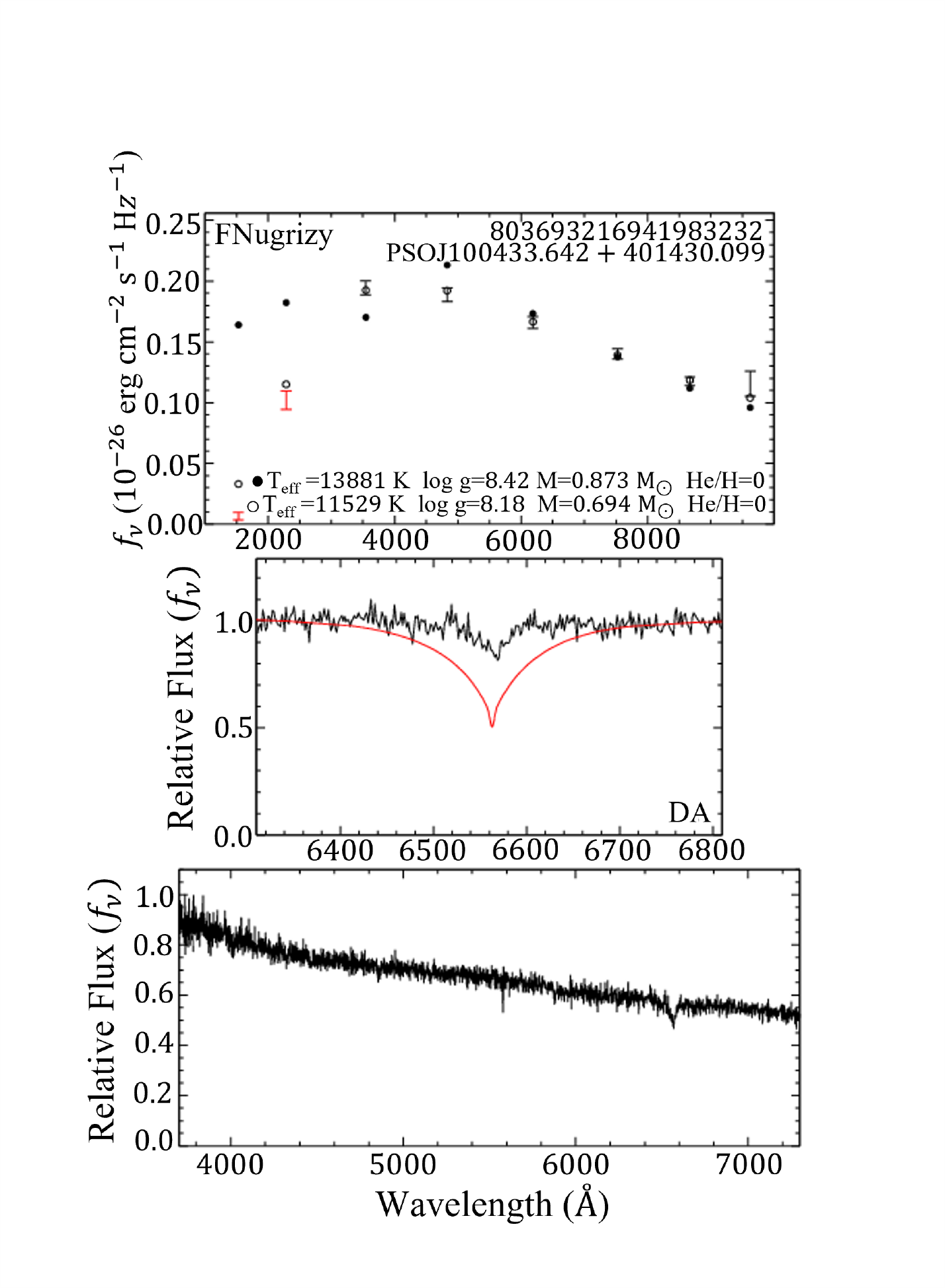}
\includegraphics[width=2.3in]{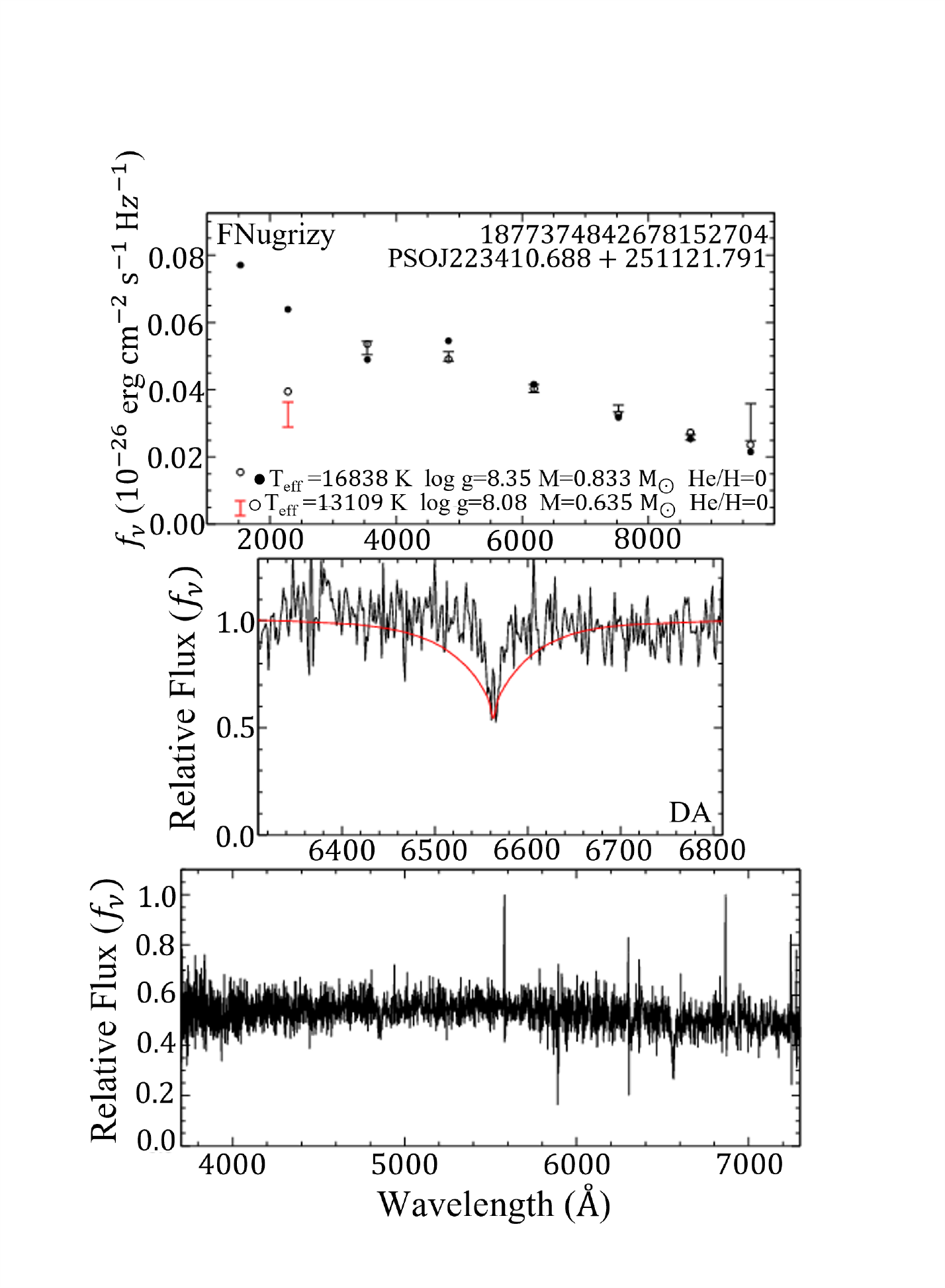}
\caption{Model atmosphere fits to three DA white dwarfs with UV flux deficits. The top panels show the best-fitting H (filled dots) and He (open circles) atmosphere white dwarf models to the optical photometry (black error bars). The middle panels show the observed spectrum (black line) along with the predicted spectrum (red line) based on the pure H atmosphere solution. The bottom panels show a broader wavelength range. GALEX FUV and NUV data clearly favor the He-dominated atmosphere solutions, which are also confirmed by the relatively weak Balmer lines in their spectra.}
\label{figberg}
\end{figure*}

\begin{figure*}
\centering
\includegraphics[width=3in]{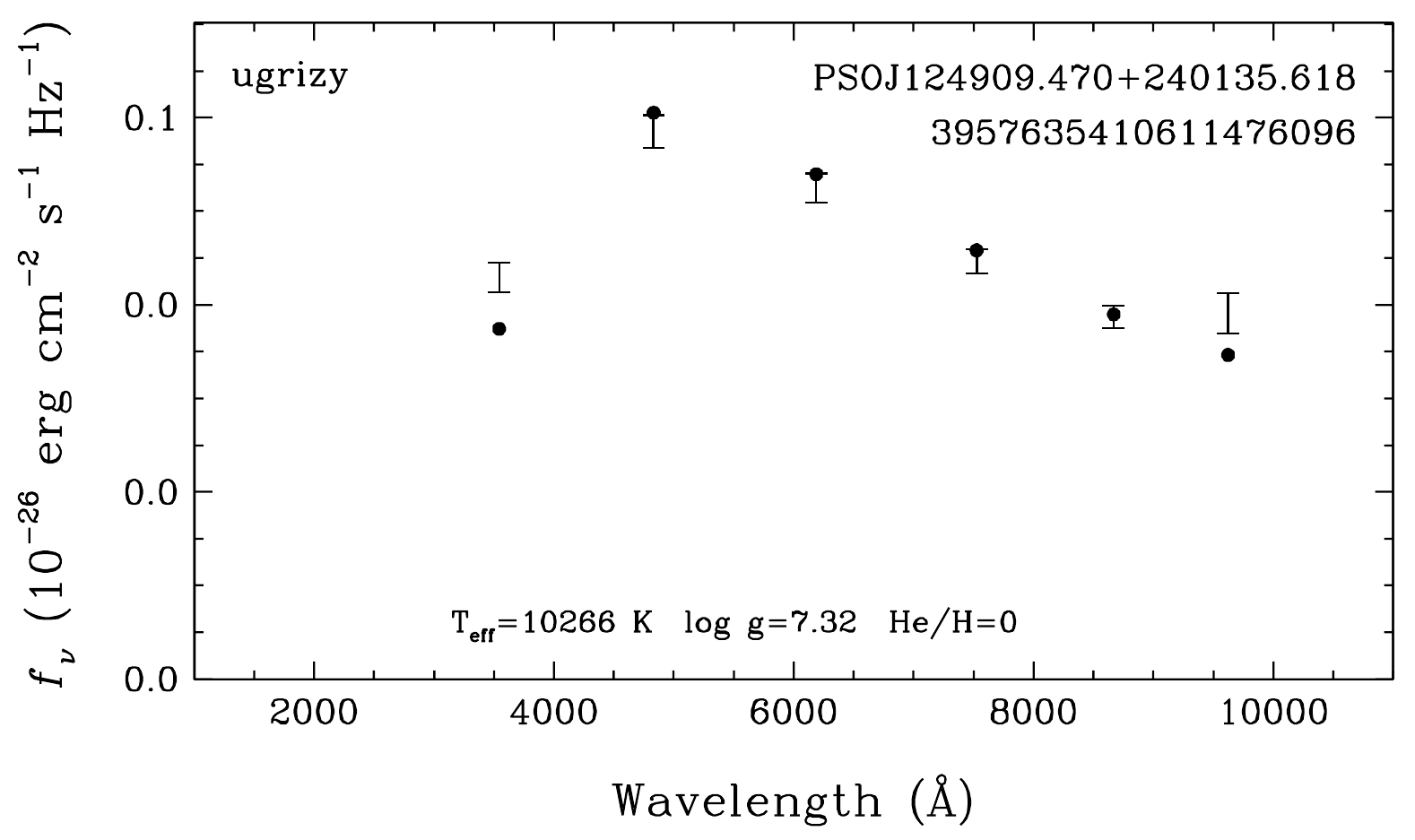}
\includegraphics[width=3in]{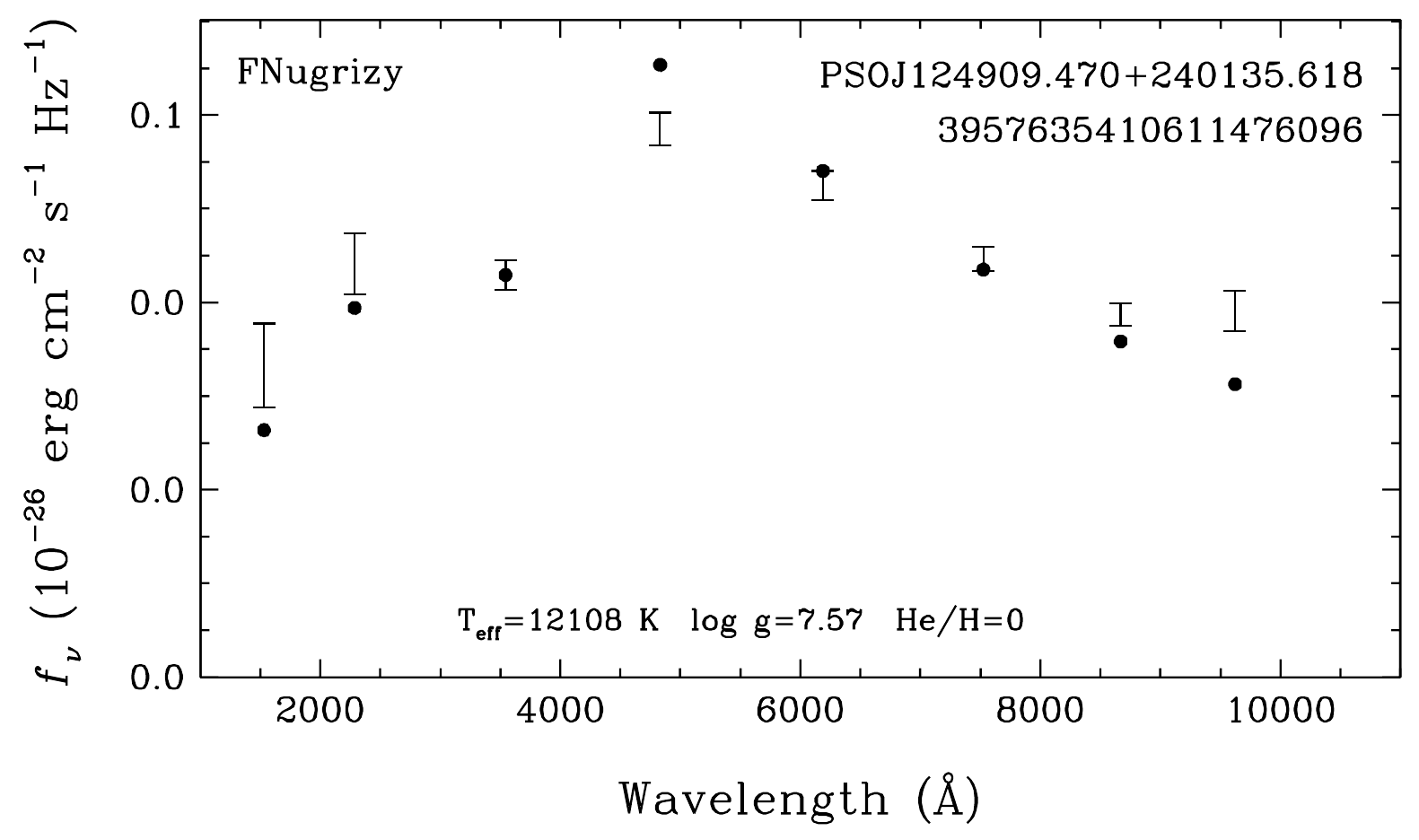}
\includegraphics[width=3in]{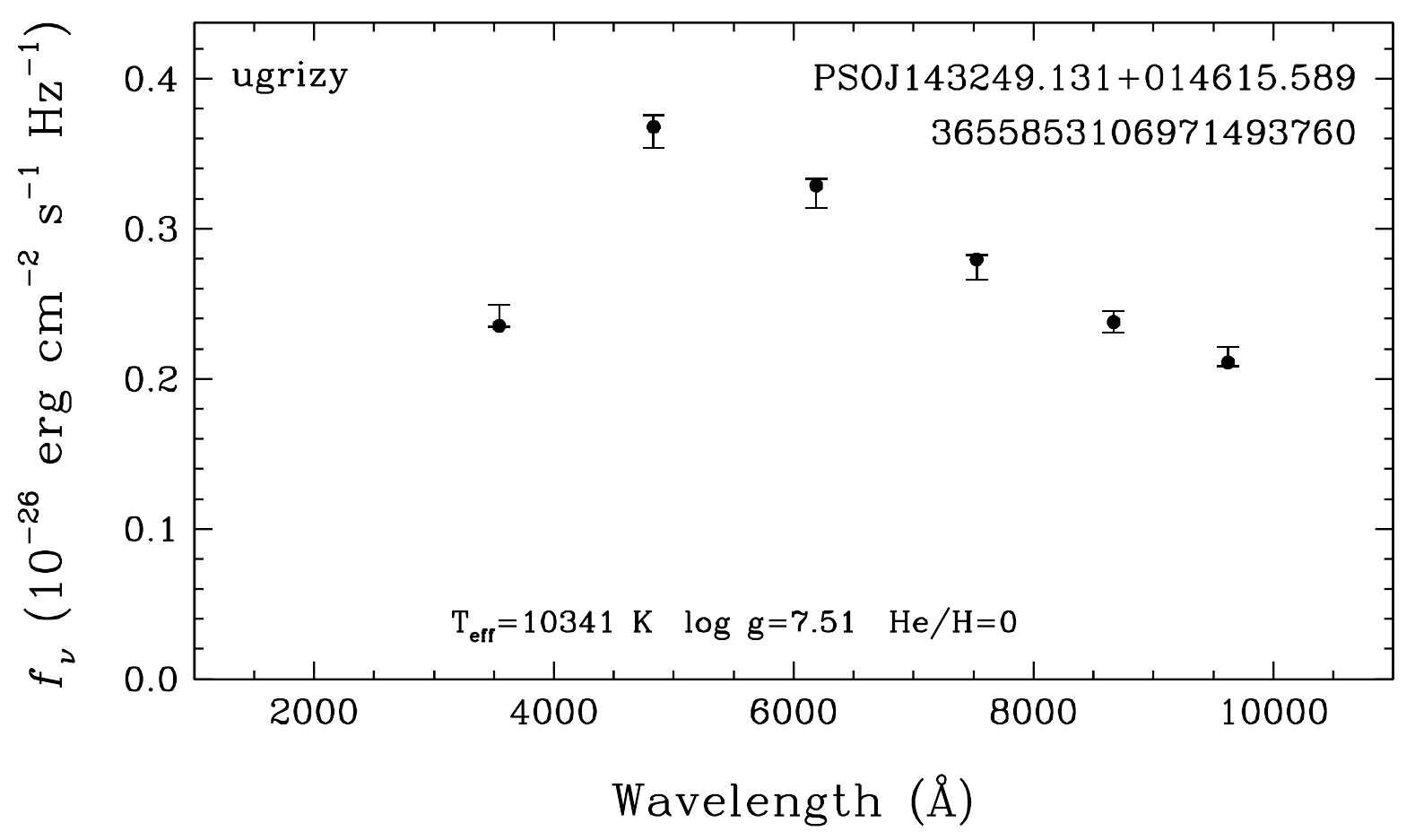}
\includegraphics[width=3in]{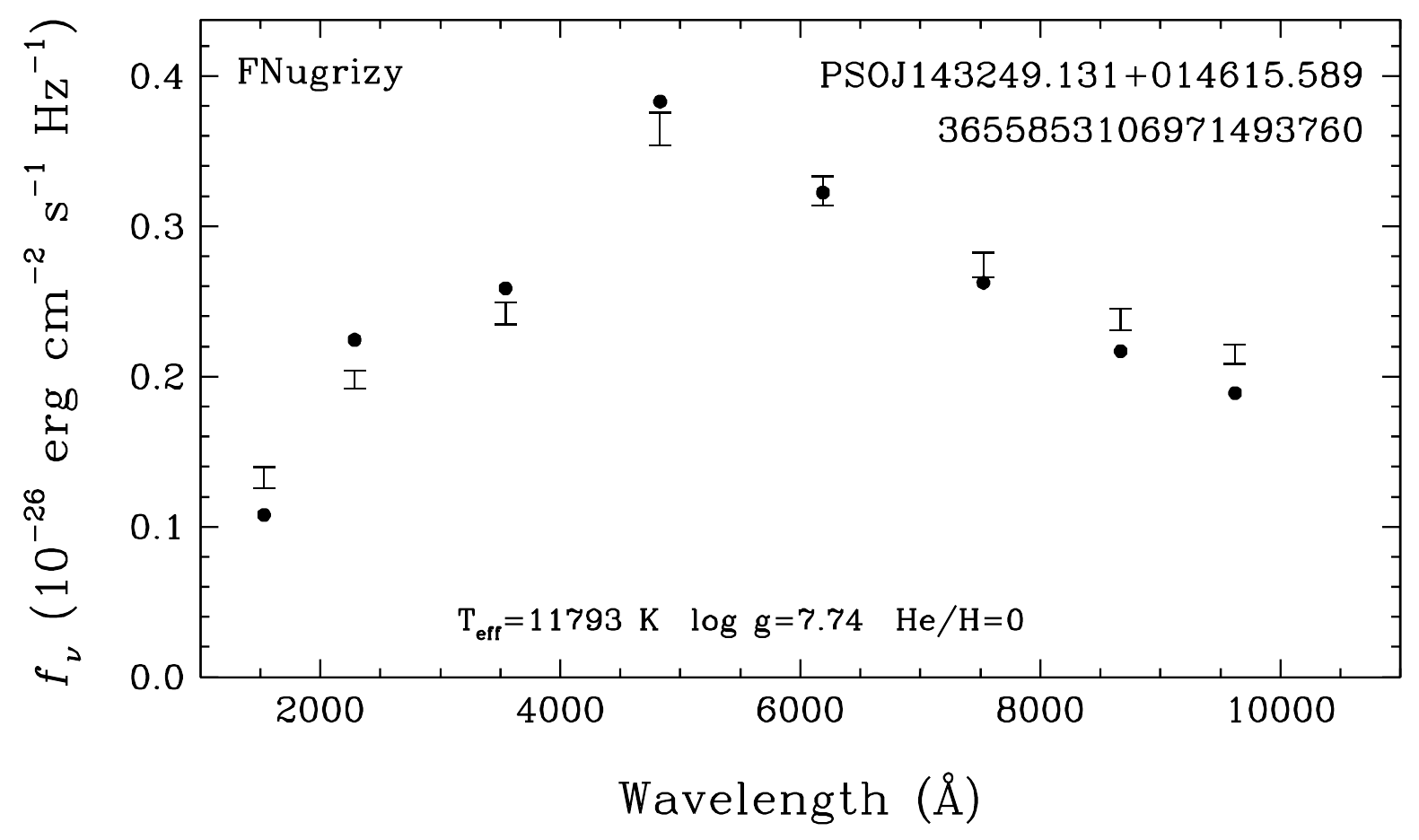}
\includegraphics[width=3in]{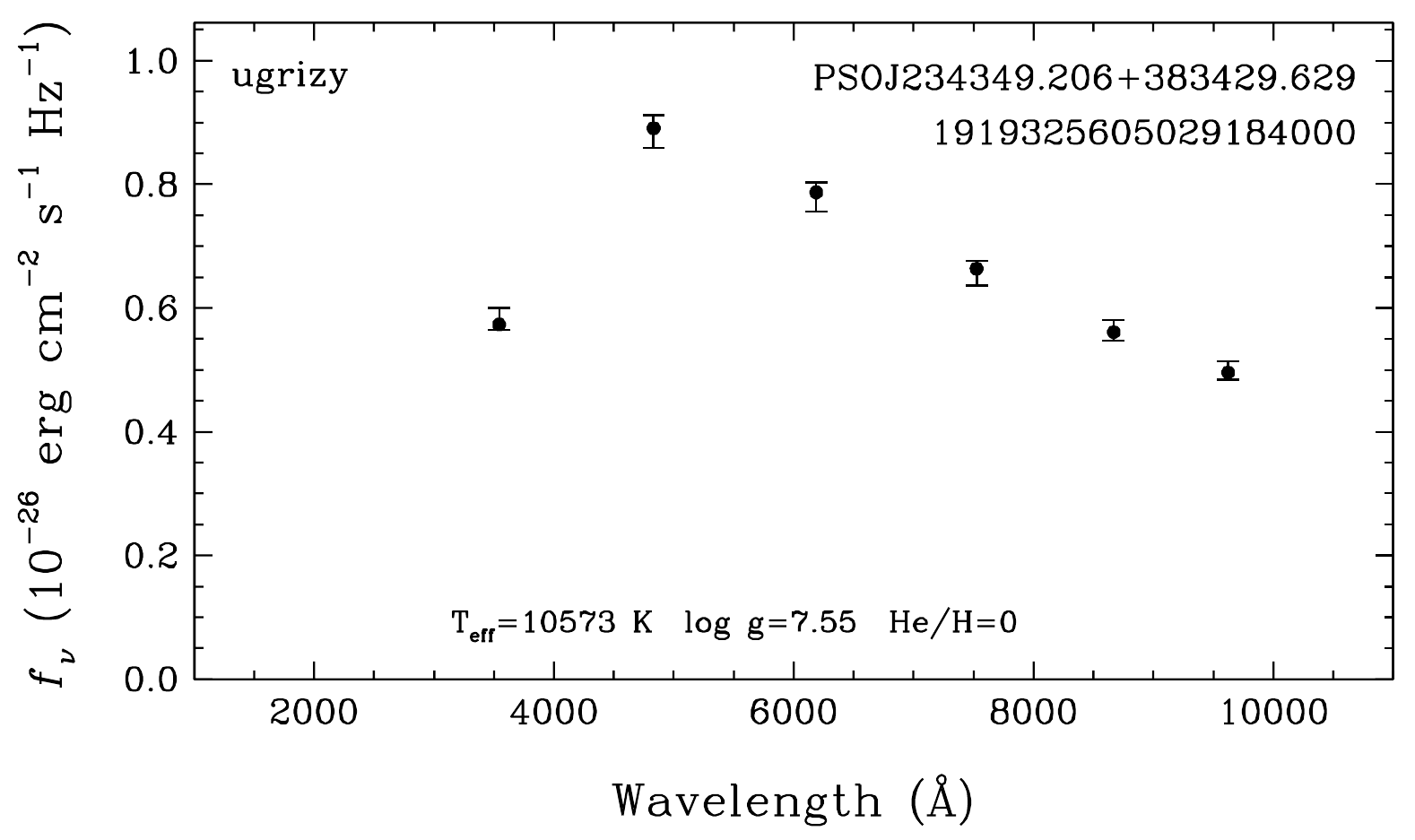}
\includegraphics[width=3in]{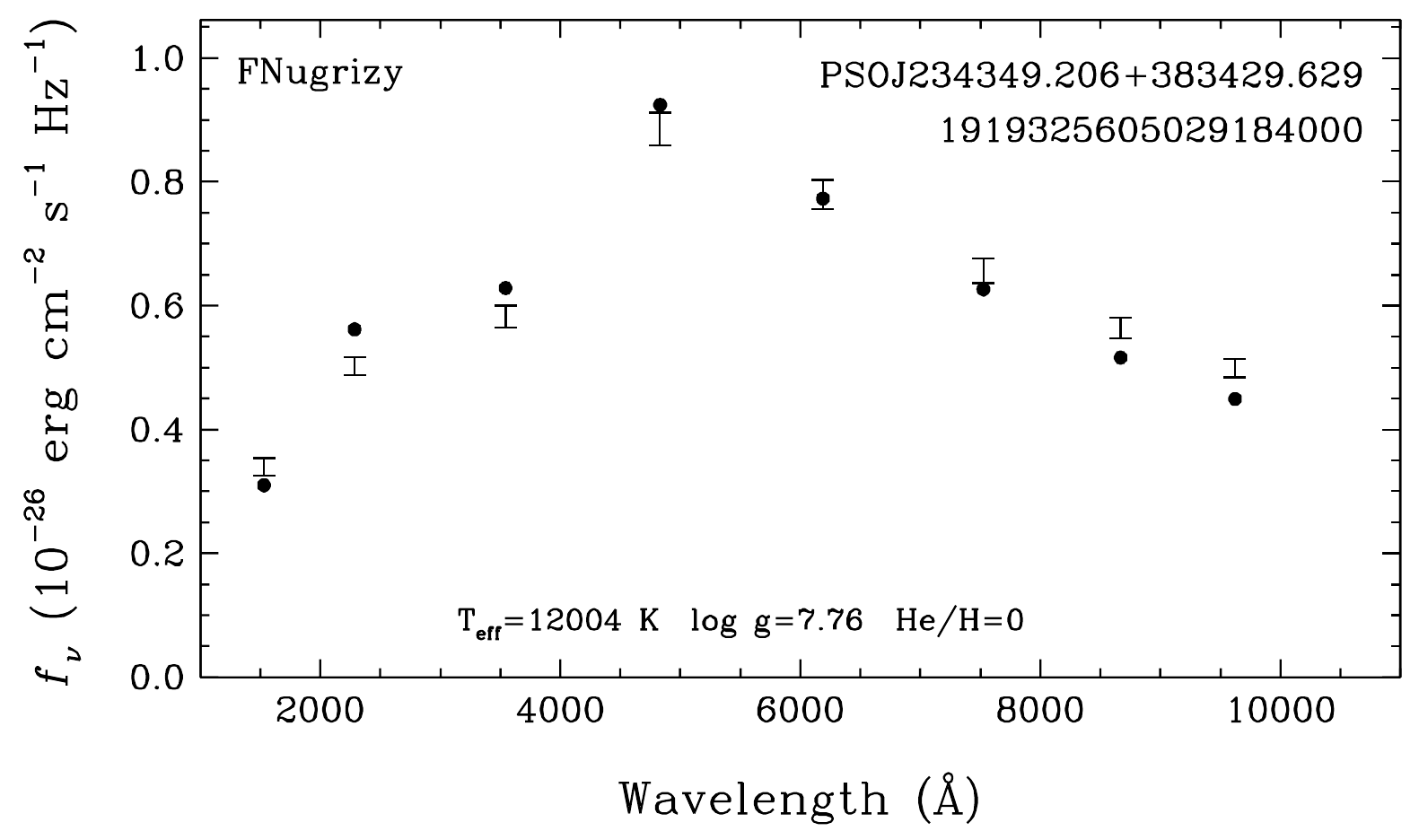}
\caption{Fits to the optical (left) and optical + UV (right) spectral energy distributions of three of the newly identified
UV excess sources in our DA white dwarf sample. The inconsistent temperature estimates from the optical and UV
photometry and optical spectroscopy indicate that they may be double white dwarfs.}
\label{fignew}
\end{figure*}

Table \ref{outnew} presents the list of 59 newly identified outliers among the DA white dwarfs with GALEX data; 24 of them
show flux deficits in the UV (their optical + UV temperatures are lower than the temperatures based on the optical data only),
and 35 are UV-excess objects. We include the spectral types from the literature for each source. 

Even though the 24 UV-deficit objects (shown in the top half of the table) are classified as DA in the literature, our analysis
indicates that they are unusual. For example, re-inspecting the SDSS spectra for three of the sources classified as DAZ
in the literature, we find that the Ca H and K lines are actually stronger than the Balmer lines, indicating that they are in fact DZA white dwarfs.

Similarly, re-inspecting the SDSS and LAMOST spectra for four of these sources
(PSO J018.6848+35.4095, J151.1401+40.2417,  J196.7725+49.1045, and J338.5445+25.1894),
we find that their Balmer lines are much weaker than expected for these relatively warm white dwarfs with $T_{\rm eff}>10,000$ K.
Figure \ref{figberg} shows the model fits to three of these objects based on the optical photometry.
All three stars are significantly fainter than expected in the FUV and NUV bands compared to the pure H atmosphere models.
The UV photometry and the weak Balmer lines indicate that
these stars are likely DA(He) white dwarfs with helium dominated atmospheres. 

The newly identified UV excess sample likely includes many binaries, including white dwarf + main-sequence and
double white dwarf systems. We classify 14 of these systems as likely DA + M dwarfs based on their spectral energy distributions,
which are dominated by the white dwarf in the UV and by a redder source in the Pan-STARRS $zy$ bands. Four of
these DA + M dwarf systems are also resolved
in the Pan-STARRS $zy$ band stacked images, but the resolved companions are not included in the Pan-STARRS photometric catalog.
However, one of these resolved systems is confirmed to be a physical binary through Gaia astrometry.
Both components of PSO J211.4189+74.6498 are detected in Gaia with source IDs Gaia DR3
1712016196599965312 and 1712016196599171840.

Figure \ref{fignew} shows the fits to the optical and optical + UV spectral energy distributions for three of the newly
identified UV excess sources that may be double white dwarfs. There are small but significant temperature discrepancies
between the photometric solutions relying on optical and optical + UV data and also the optical spectroscopy.
For example, for PSO J218.2047+01.7710 the model fits to the optical photometry give $T_{\rm eff} = 10341 \pm 329$ K and
$\log{g} = 7.51 \pm 0.05$, while the fits to the optical + UV photometry give $T_{\rm eff} =  11793 \pm 98$ K and
$\log{g} = 7.74 \pm 0.02$. Fitting the normalized Balmer line profiles, \citet{tremblay11} obtained $T_{\rm eff} =  11360 \pm 120$ K
and $\log{g}= 8.19 \pm 0.06$ for the same star. The inconsistent $\log{g}$ estimates can be explained if the photometry
is contaminated by a companion \citep[see also][]{bedard17}, and the small temperature differences between the different
solutions favor a white dwarf companion rather than a cool, late-type M dwarf star. Follow-up spectroscopy of these three
systems, as well as the rest of the UV excess sample would be helpful for constraining the nature of these objects and
identifying additional double white dwarf binaries.

\section{Results from UV Magnitude Comparison}

The optical/UV temperature comparison method presented in the previous section provides an excellent method to identify
sources with grossly different temperatures. However, it may miss some sources with unusual UV fluxes.
Those model fits rely on three ($gri$) to six ($ugrizy$) optical filters versus one or two GALEX UV filters, hence the UV data have a
lesser weight in constraining the temperatures.

To search for additional outliers that were potentially missed by the temperature comparison method, here we use model fits to the optical photometry
plus Gaia parallaxes to predict the brightness of each star in the GALEX filters, and search for significant outliers using FUV and NUV data.
To obtain the best constraints on the predicted FUV and NUV brightnesses of each source, we further require our stars to have photometry in the SDSS $u$ filter as well as all of the Pan-STARRS filters. Our final magnitude comparison sample contains 10049 DA white dwarfs with photometry in at least one of the GALEX filters, the SDSS $u$, and the Pan-STARRS $grizy$ filters.

\begin{figure*}
\includegraphics[width=3.4in, clip=true, trim=0.5in 0.2in 0.7in 0.6in]{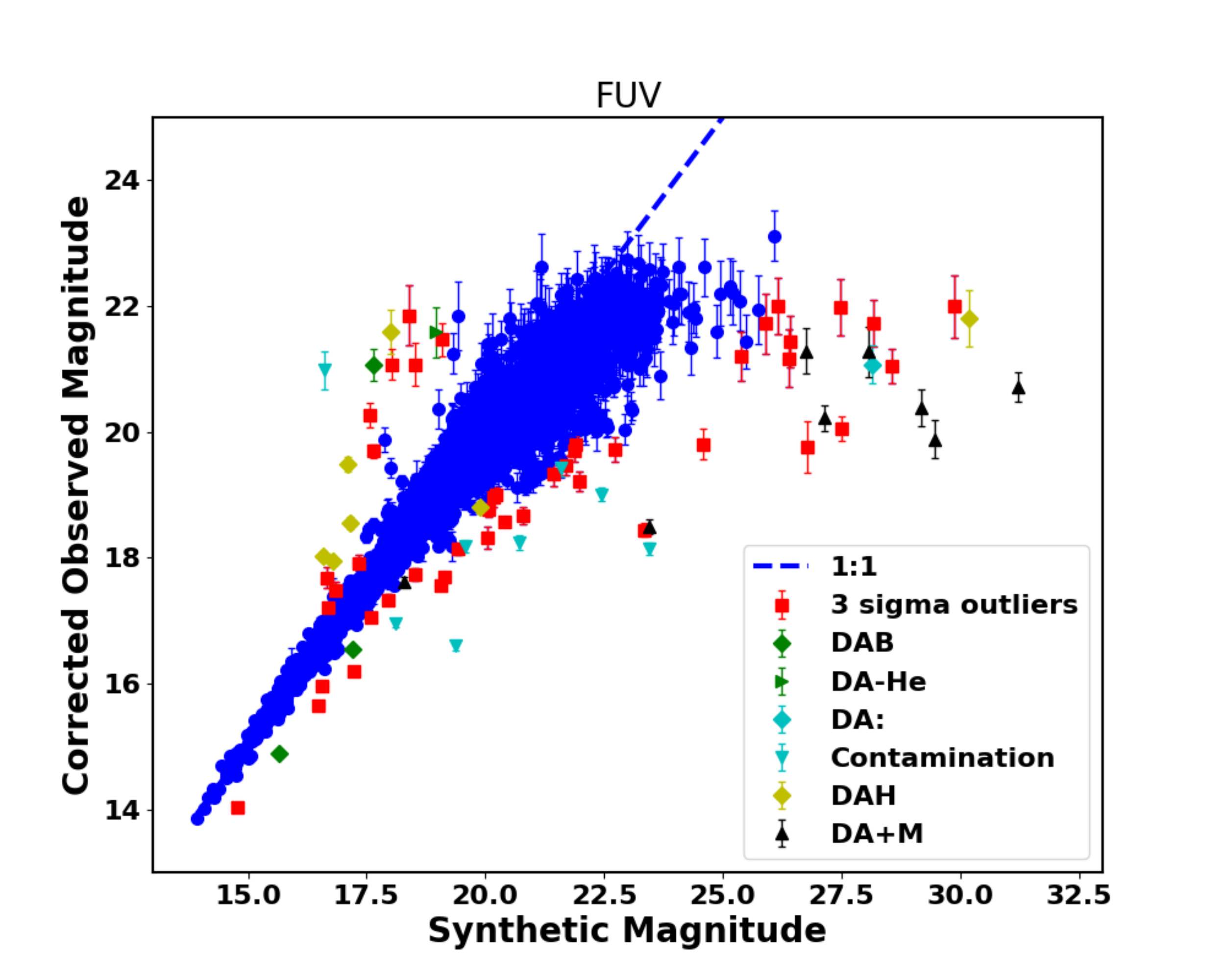}
\includegraphics[width=3.4in, clip=true, trim=0.5in 0.2in 0.7in 0.6in]{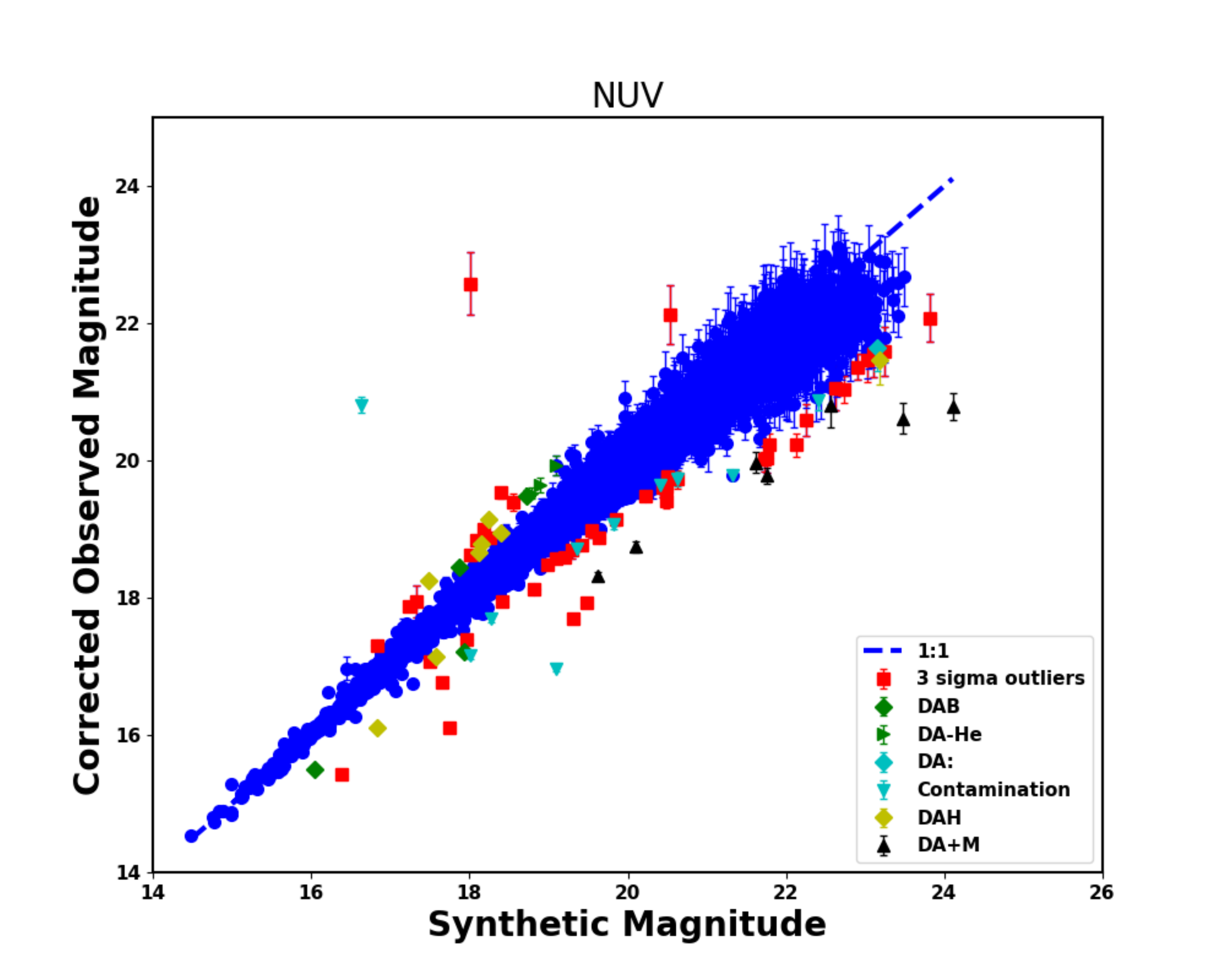}
\caption{Comparison between observed and model FUV (left) and NUV (right) magnitudes.
The blue dashed line is the 1:1 correlation. The green diamonds are previously known DAB white dwarfs,
the green triangles are previously known DA-He white dwarfs, the cyan diamonds are white dwarfs with uncertain
classifications, the cyan triangles are objects with contaminated photometry, and the yellow diamonds are previously
known magnetic white dwarfs. Previously unknown 3$\sigma$ outliers are plotted as red squares.}
\label{fig:synthvsobs}
\end{figure*}

\begin{figure*}
\centering
\includegraphics[width=4.8in, clip=true, trim=0.3in 0.1in 0.4in 0.4in]{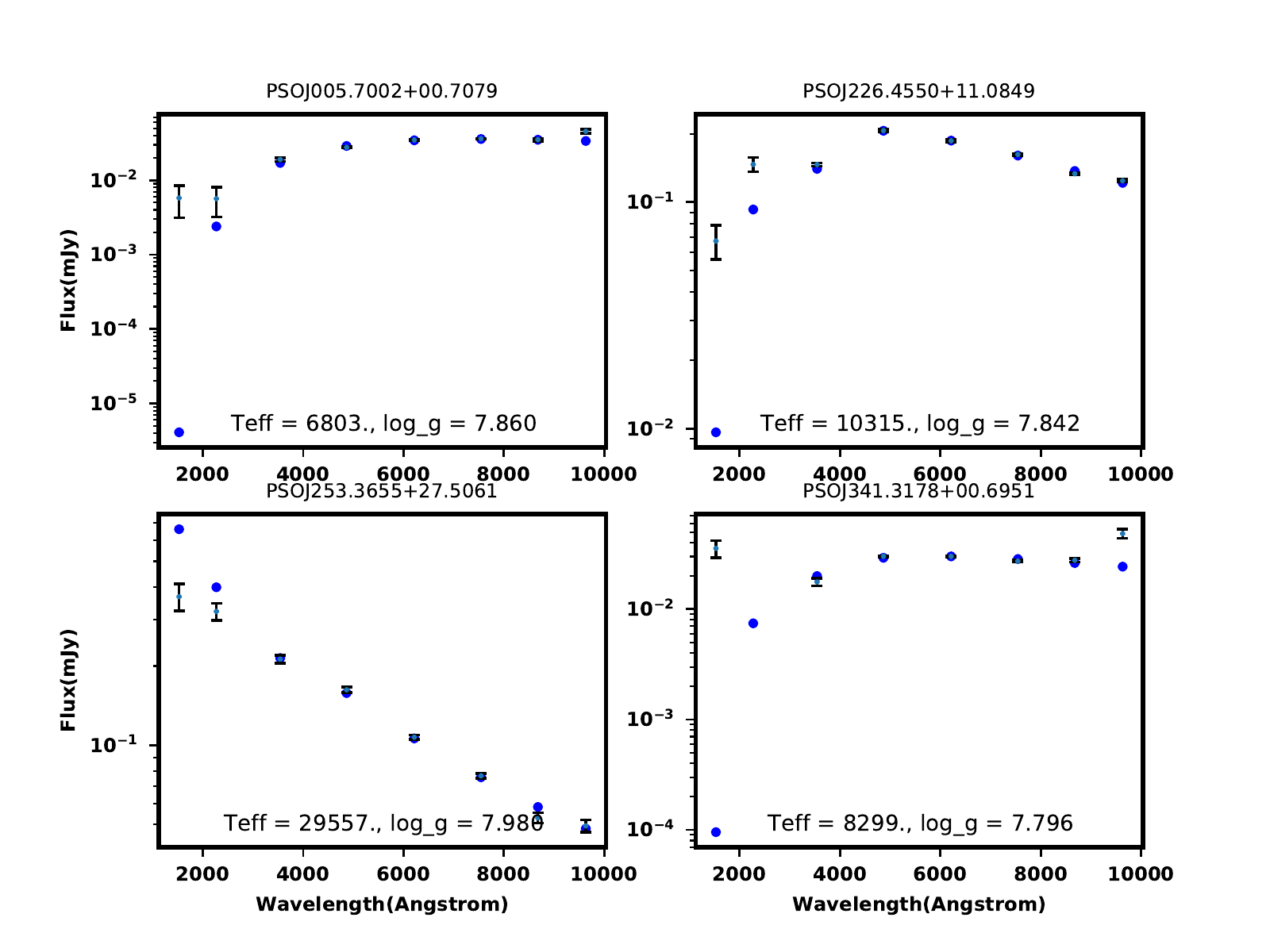}
\caption{Spectral energy distributions of four newly identified outliers in the magnitude comparison sample. The filled dots are the model fluxes and the error bars are the observed photometry.}
\label{fig:out}
\end{figure*}

Figure \ref{fig:synthvsobs} shows a comparison of the observed and predicted FUV (left) and NUV (right panel) magnitudes of the
10049 DA white dwarfs in our magnitude comparison sample. The blue dashed line is the 1:1 correlation between
observed and model magnitudes. The green diamonds are previously known DAB white dwarfs while the green triangles are DA white dwarfs that have significant amounts of helium in their atmospheres, making the use
of pure hydrogen atmosphere models inappropriate. The yellow diamonds are previously known magnetic
white dwarfs and the black triangles are previously known DA + M dwarf systems. The blue diamonds
are white dwarfs with uncertain (e.g., DA:) classifications. 

As with the temperature comparison sample, blending and contamination from background sources is an issue for some sources. We checked the Pan-STARRS stacked images for each of these outliers to identify nearby sources that could impact GALEX, SDSS, or Pan-STARRS photometry measurements. The outliers that were affected by contamination are marked by blue triangles in Figure \ref{fig:synthvsobs}.
The red squares are 30 newly identified $3\sigma$ outliers. Table \ref{magnew} presents this list along with their photometric and spectroscopic
temperatures based on the optical data.  

Figure \ref{fig:out} displays the spectral energy distributions for four of these outliers. Outliers with UV excesses, such as PSO J226.4550+11.0849
shown in the top right panel of Figure \ref{fig:out}, are likely binaries. Outliers with UV deficits, such as PSO J253.3655+27.5061 shown
in the bottom left of Figure \ref{fig:out}, do not fit the expectations from pure hydrogen atmosphere models in the UV. Their atmospheres might be dominated
by helium or might contain metals, making the use of pure hydrogen models inappropriate. Alternatively, they could also be magnetic. Further observations are needed to confirm the nature of these UV excess and UV deficit objects.

\begin{table*}
\centering
\caption{Additional outliers identified through a comparison of the observed and predicted UV magnitudes.}
\begin{tabular}{crrrlll}
\hline
Object  & Gaia DR3 Source ID &  Photometric  & Spectroscopic  & Spectral & Reference & Notes\\
  &  & $T_{\rm eff}$ (K) & $T_{\rm eff}$ (K) & Type & & \\
\hline
PSO J001.0830+23.8334 & 2849729771768028544  & 27453 & 34738 & DA & \citet{kepler16} &   \\
PSO J004.9372+33.6842 & 2864011530163554816  &  7513 &  8982 & DA & \citet{kepler16} &   \\
PSO J005.7002+00.7079 & 2546893650655427840  &  6803 &  6992 & DA & \citet{kleinman13} &  DAM? \\
PSO J021.8549+27.6214 & 296372465914661248   &  6823 &  6723 & DA & \citet{kepler16} &   \\
PSO J056.0308-05.2121 & 3244802712151826048  & 10331 & 12371 & DA & \citet{kleinman13} &  DAM? \\
PSO J118.9063+21.1283 & 673549759340742272   &  9270 &  9941 & DA & \citet{kleinman13} &  DAM? \\
PSO J126.3419+17.4310 & 662102679359467648   &  7867 &  7838 & DA & \citet{kleinman13} &   \\
PSO J137.9380+35.5266 & 714377928911156992   & 14027 & 19527 & DA & \citet{kleinman13} &   \\
PSO J149.4951+57.6078 & 1046386971133757184  & 10292 & 11288 & DA & \citet{kleinman13} &  DAM? \\
PSO J152.4806+00.1622 & 3831830527112439936  & 10569 & 10513 & DA & \citet{kleinman13} &  \\
PSO J176.3500+24.1592 & 4004972723377902592  &  7188 &  7403 & DA & \citet{kepler16} &   \\
PSO J189.9978+33.1080 & 1514768341766532992  & 10348 & 10957 & DA & \citet{kleinman13} & DAM? \\
PSO J204.9827+60.1751 & 1662524184641472640  &  7888 &  9463 & DA & \citet{kleinman13} &  \\
PSO J205.8897+23.2339 & 1443624343108905216  &  9516 & 10373 & DA & \citet{kleinman13} &  \\
PSO J210.9347+37.1660 & 1483513830393895680  &  8703 & 11040 & DA & \citet{kepler15} & DAM? \\
PSO J213.9910+62.5129 & 1666750569898974208  &  9593 & 10114 & DA & \citet{kleinman13} &  \\
PSO J226.4550+11.0849 & 1180520345976350208  & 10315 & 11354 & DA & \citet{kleinman13} &  \\
PSO J226.6089+06.6459 & 1160300056558791168  &  9500 & 10670 & DA & \citet{farihi12} &  \\
PSO J227.2923+37.1129 & 1292306146987734784  &  8264 &  8526 & DA & \citet{kepler15} &  \\
PSO J244.4451+40.3379 & 1380686815769537920  &  7600 & 13013 & DA & \citet{kepler15} & DAM? \\
PSO J248.9274+26.3827 & 1304383217063475968  & 30346 & 34544 & DA & \citet{kleinman13} &  \\
PSO J249.2986+12.8853 & 4459617994029737216  &  7824 &  7904 & DA & \citet{kepler15} & DAM? \\
PSO J250.5693+22.9411 & 1299405148103896832  & 11188 & 12763 & DA & \citet{kleinman13} &   \\
PSO J251.3785+41.0348 & 1356243233471452288  &  7884 &  8068 & DA & \citet{kepler15} &  \\
PSO J253.3655+27.5061 & 1306991499163308160  & 29557 & 30472 & DA & \citet{kleinman13} &  \\
PSO J328.6059-00.6697 & 2680152673235328768  & 17608 & 20257 & DA & \citet{kleinman13} & \\
PSO J331.0859+24.2120 & 1795394701659196032  &  6847 &  6873 & DA & \citet{kepler15} & DAM? \\
PSO J341.3178+00.6951 & 2653703714870987648  &  8299 &  9611 & DA & \citet{kleinman13} & DAM? \\
PSO J349.8567+07.6224 & 2664938112366990080  &  7394 &  8519 & DA & \citet{kepler16} &  \\
PSO J358.8416+16.8000 & 2773308246143281920  &  7149 &  7066 & DA & \citet{kepler16} &  \\
\hline
\label{magnew}
\end{tabular}
\end{table*}

\section{Conclusions}

We analyzed the UV to optical spectral energy distributions of 14001 DA white dwarfs from the Montreal White Dwarf Database, taking advantage
of the GALEX FUV and NUV data and Gaia DR3 parallaxes. Using the 100 pc sample where extinction is negligible, we demonstrated
that there are no major systematic differences between the best-fit parameters derived from optical only data and the optical + UV photometry.
The effective temperatures derived from optical and UV + optical data differ by only $50^{+215}_{-71}$ K.
The addition of GALEX FUV and NUV data in the model atmosphere analysis helps improve the statistical errors in the fits, especially for hot
white dwarfs. 

We used two different methods to identify UV excess or UV deficit objects. In the first method, we compared the temperatures obtained
from fitting the optical data with those obtained from fitting optical + UV data. We identified 111 significant outliers with this method, including
52 outliers that were previously known to be unusual. These include DA white dwarfs with helium dominated atmospheres, magnetic white
dwarfs, double white dwarfs, and white dwarf + M dwarf systems. Out of the 59 newly identified systems, 35 are UV excess and 24 are UV deficit objects. 
In the second method, we used the optical photometry to predict the FUV and NUV magnitudes for each source, and classified sources with
$3\sigma$ discrepant FUV and/or NUV photometry as outliers. Using this method, we identified 30 additional outliers.

Combining these two methods, our final sample includes 89 newly identified outliers. The nature of these outliers cannot be constrained
by our analysis alone. Many of the UV excess objects are likely binaries, including double degenerates and white dwarfs with late-type stellar
companions. Follow-up spectroscopy and infrared observations of these outliers would help constrain their nature.

There are several current and upcoming surveys that are specifically targeting large numbers of white dwarfs spectroscopically.
For example, the Dark Energy Spectroscopic Instrument Data Release 1 is expected to contain spectra for over 47000 white dwarf
candidates \citep{manser23}. DA white dwarfs make up the majority of the white dwarf population. Hence, the number of spectroscopically
confirmed DA white dwarfs will increase significantly in the near future. The Ultraviolet Transient Astronomy Satellite
\citep[ULTRASAT,][]{benami22} will perform an all-sky survey during the first 6 months of the mission to a limiting magnitude of 23
to 23.5 in its 230-290 nm NUV passband. This survey will be about an order of magnitude deeper than GALEX. 
Future analysis of these larger DA white dwarf samples with GALEX FUV/NUV or ULTRASAT NUV data would provide
an excellent opportunity to identify unusual objects among the DA white dwarf population.

\section*{Acknowledgements}

This work is supported by NASA under grant 80NSSC22K0479, the NSERC Canada,
the Fund FRQ-NT (Qu\'ebec), and by NSF under grants AST-1906379 and AST-2205736.
The Apache Point Observatory 3.5-meter telescope is owned and operated by the Astrophysical Research Consortium.

This work has made use of data from the European Space Agency (ESA) mission
{\it Gaia} (\url{https://www.cosmos.esa.int/gaia}), processed by the {\it Gaia}
Data Processing and Analysis Consortium (DPAC,
\url{https://www.cosmos.esa.int/web/gaia/dpac/consortium}). Funding for the DPAC
has been provided by national institutions, in particular the institutions
participating in the {\it Gaia} Multilateral Agreement.

\section*{Data availability}

The data underlying this article are available in the MWDD at
http://www.montrealwhitedwarfdatabase.org and also
from the corresponding author upon reasonable request.

\input{renae.bbl}

\bsp
\label{lastpage}

\end{document}